# Pricing Catastrophe: How Extreme Political Shocks Reprice Sovereign Risk, Beliefs, and Growth Expectations[1]

Riste Ichev    Rok Spruk

## Abstract

*Extreme political shocks may reshape economies not only through contemporaneous disruption but by altering beliefs about the distribution of future states. We study how such belief ruptures affect the cost of capital, expectations, and macroeconomic dynamics, using the October 7, 2023 attack on Israel as a precisely timed shock. Leveraging monthly data from 2008 to 2025 and a donor pool of advanced economies, we estimate counterfactual paths using a matrix completion design with rolling-window cross-validation and placebo-based inference, corroborated by synthetic difference-in-differences. We document three core findings. First, Israel's long-horizon sovereign risk is persistently repriced. Ten-year yields and spreads relative to the United States rise sharply and remain elevated. Second, household welfare beliefs deteriorate durably, as reflected in consumer confidence. Third, medium-run momentum improves, captured by a strong rise in the OECD composite leading indicator. These patterns reveal risk-growth decoupling where tail-risk premia rise even as medium-horizon activity expectations strengthen. Our results highlight belief-driven channels as a central mechanism through which extreme ruptures shape macro-financial outcomes.*

---

[1] Ichev: Assistant Professor of Economics, School of Economics and Business, University of Ljubljana. E: riste.ichev@ef.uni-lj.si. Spruk (corresponding author): Associate Professor of Economics, School of Economics and Business, University of Ljubljana. E: rok.spruk@ef.uni-lj.si

# 1 Introduction

Financial markets are forward-looking institutions. They embed beliefs not only about expected cash flows, but about the entire future distribution of economic outcomes. This distinction becomes critical when political events generate discontinuities, namely, shocks that do not merely shift expected growth, but alter the perceived probability of rare and extreme outcomes. A large body of theoretical work shows that small changes in disaster probabilities or tail risk can have first-order effects on asset prices, risk premia, and long-horizon interest rates (Rietz 1988, Barro 2006, Barro and Ursúa 2008, Gabaix 2012, Wachter 2013, Farhi and Gabaix 2016, Nakamura et al. 2013). Yet, despite the centrality of these ideas in modern macro-finance literature, there is remarkably little causal evidence on how extreme political shocks, rare, sharply dated events that reconfigure expectations about security, fiscal capacity, and institutional stability, translate into the pricing of long-term risk and the formation of economic expectations.

This paper studies how markets and expectations respond to an extreme geopolitical rupture. The analysis exploits the sharply dated tragic events of October 7, 2023, which constitute a sudden, large-scale political shock with profound implications for security, fiscal policy, and institutional credibility. The objective is not to document that asset prices moved, since this is tautological *per se*, but to trace how such a rupture is capitalized into (i) the sovereign risk premium and the cost of capital, and (ii) the expectations and leading indicators that connect risk pricing to real economic adjustment. Conceptually, the paper asks a simple but underexplored question. Do extreme political shocks generate only transient turmoil, or do they induce persistent revisions of beliefs about tail risk and medium-run economic prospects?

Theoretical work provides strong reasons to expect persistence. In rare-disaster frameworks, political upheavals act as informational shocks that reveal latent vulnerabilities, thereby shifting beliefs about the probability of catastrophic states and generating durable movements in risk premia even if expected growth stabilizes (Barro 2006, Gabaix 2012, Wachter 2013, Gourio 2012). A parallel literature emphasizes that political uncertainty operates as a distinct priced risk factor, affecting discount rates and valuations independently of expected cash flows (Pástor and Veronesi 2012, Kelly et. al. 2016; Pástor and Veronesi 2013). In these models, political shocks need not permanently reduce expected output to generate persistent asset-price effects. It is sufficient that they alter the distribution of outcomes or the credibility of future policy.

At the same time, a large empirical literature shows that uncertainty shocks affect real activity through real-option, precautionary saving, and expectation-formation channels (Bloom 2009, Bachmann et. al. 2013, Fernández-Villaverde et al. 2015). More recent work emphasizes that uncertainty is not a single object. Realized volatility, latent uncertainty, forecast dispersion, and subjective expectations respond differently to shocks (Jurado et. al. 2015, Coibion et. al. 2022). This heterogeneity implies that the same political rupture may have sharply divergent effects across households, firms, and financial markets, and across horizons. Yet most existing studies focus on either asset prices or real outcomes in isolation, leaving the joint dynamics of pricing and expectations largely unexplored.

Geopolitical risk represents a particularly salient form of political uncertainty. Measures of geopolitical risk correlate strongly with global macroeconomic conditions and asset prices (Caldara and Iacoviello 2022), and episodes of international conflict are associated with heightened volatility and capital reallocation (Bekaert et. al. 2013). However, the bulk of this literature relies on aggregate indices or broad cross-country correlations, which complicate causal interpretation. Moreover, many political events unfold gradually or are anticipated by markets, making it difficult to isolate belief updating in a narrow window. In contrast, extreme geopolitical ruptures - rare, abrupt, and unanticipated - provide a uniquely powerful laboratory for studying how markets and expectations adjust when political risk becomes salient in a first-order way.

Empirically, however, such events pose a fundamental challenge. Foremost, there is only one treated unit, and the shock coincides with global macro-financial forces. Conventional event studies are well suited for high-frequency equity returns, but less informative for lower-frequency macro andسovereign outcomes. Standard difference-in-differences designs struggle when treatment is unique, time-varying unobservables are important, and the treated unit's exposure to global shocks may itself change after treatment. This paper therefore adopts a modern panel-level counterfactual approach designed for precisely this environment.

The empirical strategy is to construct the counterfactual evolution of outcomes in the absence of October 7 using a donor pool of financially integrated, institutionally stable economies not exposed to a similar large-scale contemporaneous geopolitical rupture. Identification relies on the ability of the donor pool, combined with a flexible latent-factor structure, to reproduce pre-treatment dynamics. The main estimates use matrix completion methods for causal analysis in panel data, which generalize synthetic control and interactive fixed effects and are well suited to settings with common global shocks and unit-specific loadings (Bai 2009, Abadie et. al. 2010; Athey et al. 2021). The design is explicitly dynamic. Treatment effects are estimated month-by-month, allowing a sharp distinction between immediate repricing and medium-run persistence. Robustness checks include synthetic difference-in-differences (Arkhangelsky et al. 2021), placebo-in-time and placebo-in-space tests, donor-pool perturbations, and sensitivity to regularization choices.

The analysis uses monthly data spanning the modern financial regime and focuses on outcomes that jointly measure risk pricing and expectations: long-term government bond yields and sovereign spreads (relative to the United States), broad financial-market indices, business tendency indicators, consumer confidence, and the OECD Composite Leading Indicator (CLI). This multi-outcome design is not merely a robustness exercise. It is essential for interpretation. Risk premia are equilibrium objects as they reflect beliefs and pricing kernels, not simply realized fundamentals. A credible empirical account must therefore triangulate price responses with the expectation measures that reveal how households and firms revise their beliefs about the future.

Three headline findings emerge. First, large-scale attack on Israel on October 7 generated a large and persistent increase in the sovereign risk premium. Long-term borrowing costs and yield spreads rise sharply relative to the counterfactual path and remain economically meaningful over the post-treatment horizon. The dynamics are inconsistent with a purely transitory volatility shock. Instead, they point to a sustained repricing of long-horizon tail risk. This finding provides rare causal evidence in support of rare-disaster and political-risk asset pricing theories (Barro 2006, Gabaix 2012, Wachter 2013; Pástor and Veronesi 2013) and complements a large literature on sovereign spreads and flight-to-safety episodes (Beber et. al. 2009, Longstaff et al. 2011, Krishnamurthy et. al. 2014).

Second, expectations respond heterogeneously across actors and horizons. Consumer confidence declines persistently relative to the counterfactual, consistent with sustained downward revisions in household welfare expectations. Business tendency indicators display a sharp collapse followed by partial recovery, indicative of immediate disruption and subsequent adaptation. Strikingly, the composite leading indicator rises relative to the counterfactual path, suggesting an improvement in forward-looking measures of medium-run growth momentum. This pattern is not contradictory. It highlights a central theme of the paper. Extreme political shocks can simultaneously increase long-horizon tail risk while reconfiguring medium-run growth expectations, for instance through fiscal mobilization, sectoral reallocation, and institutional adaptation. This horizon-dependent response is consistent with models in which political shocks affect discount rates and conditional distributions of outcomes rather than only expected growth (Pástor and Veronesi 2012, Gourio 2012).

Third, globally priced capital responds differently across margins. Venture capital investment collapses after 2022 and remains depressed, but similar declines are observed across the donor pool, reflecting a global repricing of growth capital following the end of the low-interest-rate regime. In counterfactual terms, the incremental October 7 effect on venture capital investment volumes is modest. This null is informative as it suggests that some capital-market quantities, especially those tied to global liquidity and exit markets, are governed primarily by the global financial cycle (Rey 2015, Miranda-Agrippino and Rey 2020), whereas sovereign spreads and local expectations embed country-specific tail risk.

Taken together, these results point to a unifying interpretation. Extreme political shocks generate horizon-dependent belief updating. Long-term yields embed persistent repricing of tail risk. Households revise welfare expectations downward, firms experience short-run disruption and adaptation, and forward-looking indicators reflect a reconfiguration of medium-run growth prospects. This pattern is precisely what one would expect if such shocks affect both the pricing kernel and expectations, but through different channels and with different speeds of adjustment.

This paper contributes to several literatures. First, it contributes to macro-finance literature by providing clean causal evidence on how an extreme geopolitical rupture is capitalized into long-horizon sovereign risk premia. While theory emphasizes the role of disaster risk and time-varying discount rates (Barro 2006, Gabaix 2012, Wachter 2013), empirical identification of political disaster shocks remains scarce. Second, it contributes to the political uncertainty literature by tracing how a single, well-identified shock propagates through asset prices and expectations (Bloom 2009, Baker et. al. 2016, Hassan et al. 2019). Third, it contributes methodologically by demonstrating how modern panel counterfactual methods can be deployed in macro-finance environments with strong global co-movement.

The remainder of the paper is organized as follows. Section 2 presents a conceptual framework linking extreme political shocks to risk premia and horizon-dependent belief updating. Section 3 describes the data, sample construction, and donor pool. Section 4 outlines the identification strategy. Section 5 presents the main results. Section 6 provides robustness and sensitivity analyses. Section 7 concludes.

## 2 Conceptual Framework

This section provides a conceptual structure for interpreting the empirical results and for organizing testable implications. The goal is not to introduce a fully parameterized model, but to formalize the key mechanisms through which an extreme geopolitical rupture can affect (i) the pricing of long-horizon risk in sovereign bond markets, and (ii) expectations and leading indicators that summarize the medium-run macroeconomic outlook. The framework emphasizes two ideas. First, extreme shocks operate primarily by shifting beliefs about tail risk and the pricing kernel rather than by mechanically reducing expected growth. Second, different economic actors and different classes of indicators respond at different horizons and with different adjustment speeds, generating a distinctive pattern across yields, spreads, financial indices, confidence, business sentiment, and composite leading indicators.

### 2.1 Extreme Political Shocks as Belief Shocks

Standard macroeconomic analysis treats shocks as innovations to fundamentals such as productivity, preferences, fiscal policy, or technology, that affect expected paths of output, consumption, and investment. In contrast, the defining feature of extreme political ruptures is not that they mechanically reduce expected output, but that they reveal information about the fragility of the economic and institutional environment. They are therefore best conceptualized as belief shocks. In such setting, belief shocks capture events that compress learning about low-probability, high-impact states of the world into a narrow window and thereby alter the perceived distribution of future outcomes.

This distinction is central to modern macro-finance. In rare-disaster models, asset prices are highly sensitive to changes in the probability mass of extreme states, even if expected growth remains stable (Gabaix 2012, Farhi and Gabaix 2016). Because marginal utility is highly convex in consumption, a small increase in the probability of catastrophic outcomes can generate large increases in risk premia. In these models, what matters is not only the mean of future consumption but its left tail. Extreme political shocks naturally operate on this margin: they reveal vulnerabilities in security, fiscal capacity, and institutional resilience that were previously latent or underestimated.

To fix ideas, consider a representative agent with CRRA preferences over consumption $C_t$, facing a stochastic growth process with occasional rare disasters. Let disaster events be governed by a latent political fragility state $s_t \in \{0,1\}$, where $s_t = 1$ corresponds to a regime in which the economy is vulnerable to large negative shocks to output, fiscal solvency, or political stability. Agents hold beliefs denoted as $\pi_t = Pr(s_{t+1} = 1|\mathcal{I}_t)$. An extreme geopolitical rupture updates these beliefs such that $\pi_t > \pi_{t-1}$. Importantly, this update can be persistent if the shock is interpreted as revealing structural vulnerability rather than transitory noise.

In such a setting, long-horizon asset prices depend on both expected growth and the pricing of tail risk. The stochastic discount factor can be written schematically as:

$$M_{t+1} = \beta \left(\frac{C_{t+1}}{C_t}\right)^{-\gamma} \times \Theta(s_{t+1})$$

where $\Theta(s_{t+1})$ captures the state-dependent distortion induced by disaster risk. Even if $\mathbb{E}[\Delta \log C_{t+1}]$ remains unchanged, an increase in $\pi_t$ can raise risk premia by increase the covariance between payoffs and the pricing kernel. This logic underpins the core insight of rare-disaster asset pricing, namely, that belief about low-probability event matter disproportionately. Political shocks are natural candidates for such belief updates. Unlike productivity or demand shocks, which have continuous dynamics, extreme political events such as wars, coups, constitutional breakdowns or terrorist attacks, carry direct information about the state's ability to provide security, maintain tax bases, and enforce contracts. Therefore, they speak directly to the joint distribution of future fiscal capacity, social stability and growth. Importantly, they may do so without mechanically reducing expected output in the short run. A war can generate large fiscal mobilization, technological reallocation, and reconstruction expectations even as it raises the probability of extreme adverse outcomes.

This belief-shock framing connects naturally to the literature on political risk pricing. By way of example, Pástor and Veronesi (2012, 2013) show that political events can move asset prices by altering the distribution of future policies and the credibility of commitments, thereby affecting discount rates independently of cash flows. Furthermore, Kelly et. al. (2016) document that political uncertainty commands a distinct risk premium in equity markets. In these models, political shocks need not permanently reduce expected growth to generate persistent asset-price effects. It is sufficient that they increase uncertainty about the policy regime or the probability of unfavorable political states.

An important implication of this perspective is that persistence is diagnostic. If a political rupture is interpreted as transitory noise, asset prices should revert quickly. If it is interpreted as revealing structural fragility, i.e., as an informative signal about the distribution of future states, then its effects on long-horizon pricing objects should be persistent. This is especially true for long-term sovereign bonds, whose payoffs are directly exposed to long-run fiscal and political risk. This logic also clarifies why political shocks can have asymmetric effects across variables. Variables that primarily load on the left tail of the distribution such as long-term sovereign yields and spreads, should respond strongly to belief updates about fragility. Variables that primarily load on the conditional mean, such as near-term activity or some leading indicators, may respond differently. Households, firms, and financial markets also differ in the objects they care about. Households may focus on welfare and job security; firms on production and orders, markets on discount rates and covariance with marginal utility. A single belief shock can therefore generate a pattern of responses that appears contradictory unless interpreted through a horizon- and agent-specific lens. This framing leads to the first central prediction of the paper.

**Prediction 1 (Belief-shock channel).** *An extreme political rupture generates persistent changes in long-horizon pricing objects, particularly sovereign risk premia, even if short-run activity stabilizes or recovers.*

In the empirical analysis, this prediction is tested using long-term government bond yields and yield spreads relative to a benchmark. A persistent increase in these objects is interpreted as evidence that the shock is priced as a durable shift in tail risk rather than as a transitory volatility episode.

### 2.2 Sovereign Yields, Risk Premia, and the Pricing of Political Tail Risk

Sovereign bond yields summarize market beliefs about the long-run fiscal, macroeconomic, and institutional viability of the state. Unlike short-maturity instruments, long-term sovereign bonds embed expectations about policy regimes, tax capacity, geopolitical stability, and the distribution of future shocks over decades. They are therefore natural objects through which markets price political fragility. To fix ideas, consider the yield on an n-period sovereign bond, $y_t^{(n)}$. Abstracting from micro-level structural frictions and term structure complexities, we can write this yield as the sum of three components:

$$y_t^{(n)} \approx \frac{1}{n}\sum_{j=0}^{n-1} \mathbb{E}[r_{t+j}] + TP_t^{(n)} + SR_t^{(n)}$$

where $\mathbb{E}[r_{t+j}]$ denotes the expected future path of short-term real rates, $TP_t^{(n)}$ is the term premium compensating investors for duration and interest-rate risk, and $SR_t^{(n)}$ is a sovereign-specific risk component reflecting compensation for fiscal stress, default risk, redenomination risk, or geopolitical vulnerability. In practice, these components are not separately observed. However, comparing yields to a benchmark sovereign, such as the U.S., provides a natural way to isolate the sovereign-specific risk component as follows:

$$\text{Spread}_t^{(n)} \equiv y_{t,ISR}^{(n)} - y_{t,US}^{(n)}$$

Under the maintained assumption that the benchmark is not itself subject to the same political rupture, this spread nets out a large fraction of global movements in expected short rates and global term premia. What remains is interpreted as the market's assessment of country-specific long-horizon risk.

#### 2.2.1 Political shocks as changes in the distribution of fiscal states

Standard macroeconomic fiscal analysis often models sovereign risk through expected debt dynamics of future primary surpluses, growth, and interest rates. While these objects matter, rare-disaster asset pricing emphasizes that what is priced is not merely the expected path, but the entire distribution of possible paths, especially its left tail. Therefore, extreme political shocks can alter this distribution in at least four ways. First, wars and security crises raise the expected variance of future government expenditures and taxes, even if expected growth recovers, which elevates the fiscal capacity risk. Investors must be compensated for the risk that future fiscal burdens exceed political or institutional limits. Second, sovereign bonds are claims on the state which amplifies both political risk and institutional fragility. If a political rupture reveals fragility in state capacity such as limits to tax enforcement, political polarization, or governance stress, then even a wealthy country can be perceived as riskier in tail states. Third, security crises often generate implicit or explicit contingent liabilities through reconstruction commitments, increased insurance guarantees, and bailouts. These liabilities are rarely priced in calm times but become salient after a rupture. And fourth, sovereign revenues are procyclical. If a political shock increases the covariance between negative growth shocks and fiscal stress, then sovereign bonds become more exposed to bad states, raising required returns. Against this backdrop, these channels do not require an immediate deterioration in observed fiscal variables. They operate through beliefs about what could happen. This logic connects directly to rare-disaster models of asset pricing (Wachter 2013), in which the pricing kernel is highly sensitive to low-probability states. Sovereign bonds are particularly exposed to this logic because their payoffs extend far into the future and depend on the state's long-run solvency and legitimacy.

### 2.2.2 *Persistence as a diagnostic: volatility vs repricing*

A central conceptual distinction in this paper is between volatility shocks and repricing shocks. Volatility shocks such as sudden uncertainty about short-term conditions, can generate sharp but transitory movements in yields. Liquidity freezes, panic selling, and flight-to-quality episodes often display this pattern (Beber et. al. 2009). In contrast, repricing shocks occur when markets revise beliefs about the *structure* of future risk. These shocks generate persistent effects on long-maturity yields and spreads.

More formally, let $P_t^{(n)}$ denote the price of an $n$-period bond. Under no-arbitrage,

$$P_t^{(n)} = \mathbb{E}\left[M_{t+1} M_{t+1}^{(n-1)}\right]$$

If an event changes the conditional distribution of $M_{t+1}$ (i.e. the pricing kernel) in a persistent way, then $P_t^{(n)}$ must adjust immediately and remain displaced until beliefs revert. This is why persistence is a powerful diagnostic. It reveals whether the shock is interpreted as transitory noise or as a structural update. The key empirical object is the dynamic treatment effect on long-term yields and spreads. If these effects remain elevated months after the shock, this is evidence that the shock is priced as a persistent change in the distribution of future states, not merely a transient episode of uncertainty.

### 2.2.3 Political risk as a priced factor

A growing literature shows that political risk commands a distinct risk premium in financial markets. Pástor and Veronesi (2012, 2013) model political uncertainty as time-varying risk about future policy regimes, showing that it can generate large asset-price movements even in the absence of cash-flow news. Furthermore, Kelly et. al. (2016) document that political uncertainty predicts stock returns, consistent with a priced political risk factor. While this literature focuses largely on equity markets, the logic applies a fortiori to sovereign bonds, whose payoffs are directly contingent on the political and fiscal regime. A sovereign bond is a long-dated claim on a political institution. If the political process becomes more volatile, polarized, or fragile, the bond's payoff becomes more exposed to low-probability but high-impact states. Importantly, political risk pricing does not require default in expectation. It requires only that default, restructuring, inflationary financing, or capital controls become more plausible in extreme states.

### 2.2.4 Global risk-off vs. sovereign repricing

A key challenge in empirical work is to distinguish between sovereign-specific repricing and global risk-off episodes. Global risk-off episodes such as financial crises or monetary tightening, raise term premia and shift demand toward safe assets (Longstaff et al. 2011, Krishnamurthy et.al. 2014). These episodes can widen spreads mechanically, even in the absence of country-specific news. This is why the focus of our analysis is on relative pricing (i.e. sovereign spreads relative to a benchmark), using a panel-level counterfactual approach.. The conceptual goal is to isolate the component of yield movements that is attributable to a country-specific political rupture rather than to global repricing. This distinction is central. If the shock merely coincided with a global risk-off episode, then yields should rise similarly across the donor pool. If, instead, the spreads diverge persistently from the donor pool, this is evidence of sovereign-specific belief updating.

### 2.2.5 Implications for identification

The conceptual decomposition above has direct implications for empirical identification. Because long-term sovereign yields embed both global components, expected paths of short rates and term premia, and sovereign-specific compensation for tail risk and fiscal fragility, the empirical objective is to isolate the latter. This motivates three design choices. First, the analysis emphasizes long maturities, which are most sensitive to revisions in long-horizon risk rather than short-lived liquidity disturbances. Second, it focuses on spreads relative to a benchmark safe sovereign, which helps net out a large share of common global movements in interest rates and risk appetite that would otherwise confound inference. Third, it prioritizes counterfactual-based identification over simple pre-post comparisons, particularly, global forces such as monetary tightening or risk-off episodes can move yields in many countries simultaneously, so the relevant question is whether the treated unit diverges from the trajectory implied by comparable economies subject to the same global environment. Within this framework, persistence is itself a diagnostic. A short-lived spike is consistent with transitory uncertainty or liquidity effects, whereas a sustained deviation in long-term spreads is more naturally interpreted as a durable repricing of sovereign tail risk and fiscal uncertainty induced by the political rupture.

### 2.2.6 Testable predictions

This framework yields the following sharp predictions:

**Prediction 2 (Sovereign repricing).** *An extreme political rupture generates an immediate and persistent increase in long-term sovereign spreads relative to a counterfactual path.*

**Prediction 3 (Long-horizon sensitivity).** *The effect is larger for longer maturities, which load more heavily on tail risk.*

**Prediction 4 (Divergence from global trends).** *The spread response is not mirrored one-for-one in the donor pool, indicating sovereign-specific repricing rather than global risk-off.*

These predictions discipline the interpretation of the results. If the spread rises briefly and reverts, the shock is interpreted as transitory. If it remains elevated, it is interpreted as a persistent belief shock about tail risk and fiscal fragility.

### 2.3 A Formal Framework of Horizon-Dependent Belief Updating

This section develops a parsimonious formal framework that organizes the empirical analysis and yields sharp, testable predictions. The objective is not to provide a fully micro founded dynamic general equilibrium model, but rather to formalize the minimal structure required to explain why an extreme political rupture can simultaneously generate a persistent increase in sovereign risk premia, a persistent decline in consumer confidence, short-run disruption followed by partial recovery in business sentiment, and an improvement in forward-looking composite indicators relative to a counterfactual path. The key mechanism is horizon-dependent belief updating: the shock alters beliefs about long-run tail risk, medium-run growth momentum, and short-run disruption in distinct ways.

Time is discrete, indexed by $t = 0, 1, 2, ….T$ At time $T$, the economy is hit by an extreme political rupture. The shock does not directly destroy productive capacity but instead reveals information about the fragility of the political, fiscal, and security environment. This information is summarized by a latent state variable $s_t \in \{0,1\}$, where $s_t = 1$ denotes a high-fragility regime. Agents do not observe $s_t$ directly but form beliefs about it. Let $\pi_t = \Pr(s_{\{t+1\}} = 1 \mid I_t)$ denote the perceived probability that the economy will be in a high-fragility regime in the next period, conditional on information set $I_t$.

The shock at time $T$ generates a discrete update in beliefs such that $\pi_T > \pi_{T-1}$. Importantly, the framework allows this update to be persistent. That is, $\pi_t$ follows a slow-moving process rather than reverting immediately, reflecting the idea that the shock is interpreted as revealing structural vulnerability rather than transient noise. In addition, the shock triggers a mobilization impulse $m_t \geq 0$, capturing fiscal expansion, defense procurement, public investment, and reconstruction-related activity. This mobilization impulse affects near- and medium-run real activity but does not directly reduce tail risk. The key feature of the model is that the shock can increase $\pi_t$ while simultaneously increasing $m_t$.

### 2.3.1  Sovereign bond pricing and tail risk

Consider a long-term sovereign bond that pays one unit of consumption at maturity $n$. Let $P_t^n$ denote its price at time $t$, and let $y_t^n$ denote its yield. Prices satisfy the standard Euler equation

$$P_t^n = E_t[M_{\{t+1\}} \cdot P_{\{t+1\}}^{\{n/1\}}]$$

with $P_t^0 = 1$, where $M_{\{t+1\}}$ is the stochastic discount factor.

Assume the discount factor is given by

$$\log M_{\{t+1\}} = -r_t^f - \gamma \cdot \Delta c_{\{t+1\}} - \lambda \cdot \varepsilon_{\{t+1\}}$$

where $r_t^f$ is the risk-free short rate, $\Delta c_{\{t+1\}}$ is consumption growth, $\gamma$ is the coefficient of relative risk aversion, and $\varepsilon_{\{t+1\}}$ captures priced aggregate risk. The sovereign bond is subject to political-fiscal stress risk. If $s_{\{t+1\}} = 1$, the bond's payoff is reduced by a fraction $\delta \in (0,1)$, capturing the possibility of default, restructuring, inflationary erosion, or capital controls. Thus, the payoff at maturity is:

$$\text{Payoff}_{t+n} = \begin{cases} 1 & \text{if no stress} \\ 1 - \delta & \text{if stress} \end{cases}$$

The expected payoff of the bond is therefore decreasing in the perceived probability of entering the high-fragility regime. Under standard assumptions, the yield can be decomposed approximately as

$$y_t^{(n)} \approx \frac{1}{n} \sum_{\{j=0\}}^{\{n-1\}} \mathbb{E}_t[r_{\{t+j\}}] + TP_t^{(n)} + SR_t^{(n)}$$

where $TP_t^{(n)}$ is the term premium and $SR_t^{(n)}$ is the sovereign risk premium. The sovereign risk premium is increasing in both the probability of stress and the pricing of that stress by the stochastic discount factor. In reduced form, this can be written as

$$SR_t^{(n)} = A_n \cdot \pi_t$$

where $A_n > 0$ and $A_n$ is increasing in $n$. This monotonicity in maturity captures the intuition that long-horizon claims load more heavily on tail risk. Because stress events affect long-run fiscal capacity, long-maturity bonds are more exposed to belief updates about fragility. The key implication is that an increase in $\pi_t$ generates an immediate increase in long-term yields, and this increase persists as long as $\pi_t$ remains elevated. This is a repricing effect rather than a volatility effect. If the shock merely increased short-run uncertainty without altering beliefs about long-run fragility, the effect on long-maturity yields would be short-lived. Persistence therefore becomes a diagnostic of belief updating rather than transitory panic. To isolate sovereign-specific repricing from global risk-off movements, define the sovereign spread relative to the United States as a benchmark country:

$$\text{Spread}_t^{(n)} \equiv y_{t,ISR}^{(n)} - y_{t,US}^{(n)}$$

If the benchmark is not subject to the same political rupture, global movements in term premia and risk-free rates largely cancel out. The remaining variation is attributable to country-specific belief updating about tail risk and fiscal fragility. This proposition provides the formal interpretation of the empirical results on long-term yields and spreads. A persistent increase in spreads is not evidence of short-run panic but evidence that markets have revised their beliefs about the long-run distribution of fiscal and political states.

*2.3.2   Households and welfare-risk beliefs*

Households differ from firms and financial investors in that their objective function is not profit maximization or state-contingent pricing, but rather the maximization of expected lifetime utility under uncertainty. As a result, household belief formation is fundamentally concerned with the distribution of future welfare rather than with near-term production or medium-run growth momentum. Extreme geo-political shocks are therefore likely to have particularly persistent effects on household expectations, even when real activity stabilizes or rebounds.

Let household welfare depend on consumption, employment stability, security, and access to public services. Households face not only income risk but also political and institutional risk, which can affect taxation, transfers, inflation, public goods provision, and the likelihood of forced relocation or emigration. These dimensions are often absent from standard macroeconomic models but are central to how households perceive extreme political events.

We summarize these concerns with a latent downside-risk index $u_t$. This index does not represent realized hardship; rather, it captures perceived exposure to adverse future states. We assume that perceived downside risk increases in fragility beliefs:

$$u_t = u_0 + u_1 \pi_t + u_2 z_t$$

with $u_1 > 0$ and where $z_t$ captures other perceived risks such as inflation or local security conditions and where $\pi_t = \Pr(S_{\{t+1\}} = 1 | I_t)$ denotes the perceived probability of entering a high-fragility regime, and $z_t$ captures other salient risks, such as inflation expectations, security concerns, or the perceived credibility of institutions. This formulation embeds three key ideas. First, households respond to beliefs, not realizations. Second, they respond to downside risk, not to the mean of future outcomes. Third, beliefs about political fragility enter directly into perceived welfare risk, not only through income channels.

Without the loss of generality, consumer confidence is modelled as a decreasing function of perceived downside risk:

$$CC_t = \hat{C} - \kappa \cdot u_t + v_t$$

with $\kappa > 0$.

A simple substitution yields:

$$CC_t = \hat{C} - \kappa \cdot u_0 - \kappa \cdot u_1 \cdot \pi_t - \kappa \cdot u_2 \cdot z_t + v_t$$

This formulation implies that consumer confidence is not simply a function of expected growth or short-run labor market conditions. Instead, it reflects beliefs about the distribution of future welfare, particularly the probability of extreme negative states. This distinction is crucial. A household may believe that the economy will grow due to fiscal mobilization or reconstruction, yet simultaneously believe that its own welfare is more precarious due to higher taxation, increased inflation risk, diminished institutional protection, or heightened security concerns. In such a case, growth expectations and welfare expectations move in opposite directions. The model explicitly allows for this divergence.

The persistence of confidence shocks arises naturally in this framework. If the political rupture is interpreted as revealing a permanent or semi-permanent increase in the probability of bad states, then $\pi_t$ remains elevated even if near-term uncertainty dissipates. In contrast to transitory volatility shocks, which primarily affect short-run dispersion, belief shocks about regime fragility affect the entire future distribution. This makes them inherently persistent. This mechanism is consistent with a large literature on belief formation under uncertainty. Households overweight salient, vivid, and emotionally charged events when forming expectations. Extreme political ruptures are precisely such events. They are interpreted not as noise, but as structural breaks in the data-generating process. As a result, belief updating is slow to reverse, even when objective conditions improve.

Furthermore, households are typically unable to fully insure against political and institutional risk. Unlike firms, they cannot easily relocate production, diversify across jurisdictions, or hedge through financial instruments. Their response to increased perceived fragility is therefore not reoptimization in a narrow sense, but rather precautionary behavior, pessimism, and reduced subjective security. In this sense, consumer confidence should be understood not as a noisy proxy for short-run consumption growth, but as a state variable capturing the perceived security of the household's future. It is a welfare-belief object. This interpretation leads to a key distinction between confidence and activity-based indicators. Business tendency indicators load on short-horizon operational feasibility. CLI loads on medium-run momentum. On the contrary, consumer confidence loads on welfare risk. There is no theoretical reason for these to move together after an extreme political rupture, and strong reasons why they should not. The framework therefore predicts that consumer confidence can remain persistently depressed even when business sentiment recovers and leading indicators improve. This pattern should not be interpreted as a contradiction. It is the expected outcome of horizon-dependent belief updating.

### 2.3.3 Firms, Disruption, and Adaptation

Firms differ fundamentally from households and financial investors in both their objective functions and their margins of adjustment. Whereas households primarily care about welfare and downside protection, and financial investors price long-horizon state-contingent payoffs, firms are concerned with near-term operational feasibility, input availability, demand realization, and financing conditions. As a result, their response to extreme political shocks is typically characterized by sharp short-run disruption followed by endogenous adaptation.

Let near-term real activity growth be given by:

$$\Delta y_{\{t+1\}} = \alpha \cdot m_t - \theta \cdot d_t + \varepsilon_{\{t+1\}}$$

with $\alpha > 0$ and $\theta > 0$ and where $m_t$ represents a mobilization or demand reallocation impulse associated with fiscal expansion, defense procurement, reconstruction, and state-led investment, and $d_t$ captures disruption such as labor reallocation, supply-chain breakdowns, regulatory bottlenecks, security restrictions, and coordination failures. The key feature is that $d_t$ and $m_t$ respond to the shock through different mechanisms and at different speeds. Disruption spikes on impact because firms must immediately adjust to a new regime of uncertainty, logistics constraints, labor mobilization, and risk. We model this as:

$$d_t = d_0 \cdot \mathbf{1} \cdot \{t = T\} + \rho_d \cdot d_{\{t-1\}}$$

with $0 < \rho_d < 1$ and where mobilization increases after the shock and remains persistent. This formulation captures the idea that disruption is initially large but decays as firms learn, substitute inputs, reorganize production, renegotiate contracts, and adapt to new demand patterns. The decay parameter $\rho_d$ reflects learning, coordination, and institutional support. Mobilization, by contrast, reflects policy choices and sectoral reallocation. Defense spending, emergency procurement, infrastructure investment, and reconstruction do not occur instantaneously but ramp up over time. We model this as:

$$m_t = m_0 \cdot \mathbf{1} \cdot \{t \geq T\} + \rho_m \cdot m_{\{t-1\}}$$

Crucially, $m_t$ does not offset $d_t$ immediately. There is a window in which disruption dominates. We define business tendency indicators as near-term expectations of activity:

$$BT_t = \mathbb{E}_t[\Delta y_{\{t+1\}}] = \alpha \cdot m_t - \theta \cdot d_t$$

This formulation embeds two important features. First, business sentiment is forward-looking but short-horizon. It reflects expected activity over the next quarter or year, not over decades. Second, it is endogenous as firms update expectations based on realized adaptation, policy responses, and demand reallocation. This structure generates a non-monotonic dynamic response. At the moment of the shock, $d_t$ jumps and dominates $m_t$, causing a sharp collapse in $BT_t$. As time passes, disruption decays, mobilization persists, and firms revise expectations upward. The recovery in business sentiment is therefore not evidence that the shock was harmless. It is evidence that firms adapt. This logic is deeply rooted in the literature on uncertainty, real options, and firm dynamics. Under heightened uncertainty, firms postpone irreversible investment and hiring, generating sharp short-run contractions (Bloom 2009). However, when uncertainty resolves or new demand materializes, activity rebounds. In political and military contexts, this rebound is often driven by state demand, sectoral reallocation, and emergency procurement rather than by private consumption. The model thus makes a sharp distinction between adaptation and reversion. Reversion would imply that the economy returns to the same regime as before. Adaptation implies that firms reorganize under a new regime, with different risk, financing, and demand conditions. This distinction is central for interpretation: a recovery in business sentiment does not contradict persistent increases in sovereign spreads or persistent declines in consumer confidence. These objects load on different belief objects and horizons.

### 2.3.4 Composite Leading Indicators and Risk-Growth Decoupling

Composite leading indicators are often misunderstood because they are neither measures of current output nor of welfare. Their purpose is to detect turning points in the business cycle by aggregating forward-looking components such as order books, production expectations, financial conditions, and sentiment measures. They are explicitly designed to be predictive, not descriptive. We model the composite leading indicator as:

$$CLI_t = w_1 \cdot BT_t + w_1 \cdot FC_t + w_3 \mathbb{E}_t[\Delta y_{\{t+2\}}] + \eta_t$$

with weights $w_i > 0$. This formulation makes explicit that the $CLI$ loads on three conceptually distinct objects, (i) near-term firm expectations ($BT_t$), (ii) financial conditions ($FC_t$), and (iii) medium-run growth expectations ($\mathbb{E}_t[\Delta y_{\{t+2\}}]$). Notice that the medium-run expectations are given by

$$\mathbb{E}_t[\Delta y_{\{t+2\}}] = a_m \cdot m_t - a_d \cdot \mathbb{E}_t[d_{\{t+1\}}]$$

with $a_m > 0$ and $a_d > 0$. The medium-run expectation captures the idea that firms and analysts expect the economy to benefit from mobilization and reconstruction, but remain cautious about lingering disruptions. Furthermore, financial conditions are modelled as

$$FC_t = \varphi_0 + \varphi_m \cdot m_t - \varphi_\pi \cdot \pi_t + \varphi_s \cdot S_t$$

This equation makes explicit that financial conditions are shaped by two opposing forces. On the one hand, increased fragility $\pi_t$ raises risk premia and tightens conditions. On the other hand, mobilization $m_t$ and stabilization $S_t$, policy responses, learning, international support, and institutional adaptation, improve medium-run financing conditions. This decomposition allows the composite indicator to move differently from consumer confidence and sovereign spreads. The indicator does not measure tail risk. It measures medium-run momentum relative to trend. It can therefore improve even when spreads rise and confidence falls. This phenomenon is not paradoxical. It reflects risk-growth decoupling. The same shock can increase the probability of extreme negative outcomes (for instance, by raising tail risk) while simultaneously increasing near- and medium-run activity through fiscal expansion, mobilization, and sectoral reallocation. This is historically common. Wartime economies often exhibit high growth and full employment alongside extreme tail risk. Asset markets price the risk; leading indicators reflect momentum. Thus, the expected direction of the composite indicator after extreme event should not be viewed as an anomaly but as a diagnostic of belief heterogeneity across horizons.

### 2.3.5 Venture capital and global financial cycle

Finally, the framework distinguishes between margins that are locally priced and those governed primarily by global financial conditions. Venture capital is a quintessential globally priced asset class. Aggregate venture capital investment responds strongly to global discount rates, global valuations of growth assets, and exit-market conditions. Country-specific changes in tail risk can matter, but they may be second-order relative to global repricing episodes, especially in annual data with limited post-treatment observations.

Let annual venture capital investment satisfy

$$\log VC_t = a - b\, R_t^g + c\, \log V_t^g - \chi\, \pi_t + \xi_t$$

where $R_t^g$ denotes global financial conditions and captures a global discount factor, $V_t^g$ denotes global valuations of growth and technology assets, and $\pi_t$ is local tail risk. A sufficient condition for global dominance is:

$$\left| b \cdot R_t^g \right| + \left| c \cdot \log V_t^g \right| \geq \left| \chi \cdot \pi_t \right|$$

When this holds, large global repricing of growth capital can generate declines across many countries simultaneously, producing weak incremental treatment effects for the shock-affected entity relative to a donor pool. This provides a disciplined explanation for why some capital-flow or funding aggregates may not display strong Israel-specific deviations in response to a large-scale domestic shock even when sovereign spreads and local confidence indicators respond sharply.

The framework yields five propositions that discipline the empirical analysis.

### 2.3.6 Testable propositions

**Proposition 1 (Sovereign repricing of tail risk).** *If the shock at time $T$ causes a discrete and persistent increase in fragility beliefs such that $\pi_T > \pi_{T-1}$ and $\pi_t$ remains elevated for $t \geq T$, then the sovereign risk premium satisfies $SR_t^{(n)} \approx A_n \pi_t$ with $A_n > 0$ increasing $n$. Consequently, long-maturity sovereign spreads $Spread_t^{(n)}$ rise sharply at $T$ and remain elevated relative to their counterfactual path, with larger effects at longer maturities.*

**Proposition 2 (Household welfare-belief channel).** *If consumer confidence satisfies $CC_t = C - \kappa \cdot u_t + v_t$ and a downside risk satisfies $u_t = u_0 + u_1 \pi_t + u_2 z_t$ with $u_1 > 0$, then any persistent increase in $\pi_t$ generates a persistent decline in $CC_t$ even if near-term activity indicators partially recover.*

**Proposition 3 (Panic-adaption in business tendency).** *Let business tendency trajectory satisfy $BT_t = \alpha \cdot m_t - \theta \cdot d_t$, with disruption captured by $d_t = d_0 \cdot \mathbf{1}\{t = T\} + \rho_d \cdot d_{t-1}$, $0 < \rho_d < 1$, and mobilization $m_t = m_0 \cdot \mathbf{1}\{t \geq T\} + \rho_m \cdot d_{t-1}$, $0 < \rho_m < 1$. Then, $BT_t$ drops* *sharply at T and subsequently recovers partially as disruption decays and mobilization persists, yielding a panic-adaptation dynamic.*

**Proposition 4 (Risk-growth decoupling and CLI).** *Let $CLI_t = w_1 \cdot BT_t + w_1 \cdot FC_t + w_3 \mathbb{E}_t[\Delta y_{\{t+2\}}] + \eta_t$ with $\mathbb{E}_t[\Delta y_{\{t+2\}}] = a_m m_t - a_d \mathbb{E}_t[d_{t+1}]$ and $FC_t = \varphi_0 + \varphi_m \cdot m_t - \varphi_\pi \cdot \pi_t + \varphi_s \cdot S_t$. Then, there exist parameter values such that an increase in $\pi_t$ raises sovereign spreads while an increase in $m_t$ and $S_t$ raises $CLI_t$, implying that $CLI_t$ can improve relative to counterfactual even when sovereign spreads rise and consumer confidence falls.*

**Proposition 5 (Global financial cycle dominance in venture capital).** *Let $\log VC_t = a - b R_t^g + c \log V_t^g - \chi \pi_t + \xi_t$. If global financial conditions and valuations move sufficiently such that $|b \cdot R_t^g| + |c \cdot \log V_t^g| \geq |\chi \cdot \pi_t|$, then VC investment declines broadly across countries and the incremental effect of local fragility beliefs $\pi_t$ on VC investment is weak relative to the global cycle, especially in short post-shock windows.*

## 3    Data, Outcomes and Samples

This section describes the datasets, outcome variables, and sample construction used in the empirical analysis. The econometric design leverages monthly data over the period 2008:M1-2025:M8 and a donor pool to estimate the effects of the October 7, 2023 extreme political rupture on multiple dimensions of economic and financial activity. The section also explains the rationale for each outcome measure in light of the conceptual framework in Section 2.

### *3.1    Outcomes and measurement*

Our outcomes are chosen to map directly to the theoretical channels in Section 2 such as long-horizon tail-risk repricing (i.e., sovereign yields/spreads), household welfare-risk beliefs (i.e. consumer confidence), firm-level disruption and adaptation (i.e. business tendency), medium-run momentum (i.e. CLI), and globally priced risk capital (i.e. venture capital)

#### *3.1.1    Sovereign yields and yield spreads*

The primary financial outcome is Israel's 10-year government bond yield, measured monthly from Bloomberg DataStream. We also study the yield spread relative to the United States, which nets out a large share of common global movements in risk-free rates and term premia. Bloomberg's rates and bonds market data offering provides government bond yields and spreads as standard market objects.

Let $y_{t,ISR}^{(10)}$ and $y_{t,US}^{(10)}$ denote the 10-year government bond yields for Israel and the U.S., respectively. The spread is:

$$Spread_t = y_{t,ISR}^{(10)} - y_{t,US}^{(10)}$$

This outcome corresponds to the long-horizon pricing object in Section 2 and ensures that persistent movements in long-maturity yields and spreads are interpreted as repricing of sovereign tail risk and fiscal-security fragility, consistent with modern political-risk asset-pricing frameworks.

### 3.1.2 Household welfare risk beliefs: consumer confidence

To measure household expectations, we use the OECD Consumer Confidence Index (CCI). The OECD defines the CCI as a "standardised confidence indicator providing an indication of future developments of households' consumption and saving." Consistent with our focus on high-frequency belief updating, we calculate monthly growth rates of the normalised CCI, closely aligned with the OECD's "Consumer Barometer" concept. The OECD states that the Consumer Barometer corresponds to "the monthly growth rate of the normalised consumer confidence indicator (CCI)." The OECD's methodology note, further documents that confidence indicators are computed from seasonally adjusted survey net balances and standardised for cross-country comparability. In the conceptual framework of Section 2, consumer confidence primarily loads on perceived downside risk to household welfare rather than on near-term production conditions. This aligns with the broader expectations literature emphasizing that household beliefs are economically meaningful state variables that can move differently from realized activity.

### 3.1.3 Firm expectations: business tendency

To measure firm-level expectations, we use an OECD business tendency index to capture business confidence indicator drawn from OECD short-term statistics and standardised survey-based confidence measures. The OECD documents the standardisation procedure for business and consumer confidence indicators and their construction from business tendency surveys of net balances of qualitative responses, harmonised across countries. As with consumer confidence, we compute monthly growth rates to capture changes in expectations and to avoid conflating level differences with shock responses. In our conceptual framework, these firm expectations correspond to short-horizon operational feasibility as they can display a sharp impact response (i.e., disruption) followed by partial recovery (i.e., adaptation), consistent with uncertainty-shock and real-options mechanisms in the modern macroeconomic literature (Bloom 2009).

### 3.1.4 Medium-run macroeconomic outlook: composite leading indicator

To capture medium-run cyclical momentum, we use the OECD Composite Leading Indicator (CLI). The OECD describes CLIs as designed to anticipate turning points and economic fluctuations relative to trend. This definition matters for interpretation. The CLI is not a welfare measure and not a tail-risk measure. Instead, it is a forward-looking signal of cyclical momentum around trend, precisely the quantity of interest that can move differently from sovereign spreads or consumer confidence in the risk-growth decoupling mechanism of Section 2.

### 3.1.5 Venture capital investment

Finally, we examine venture capital investment measured in million USD, obtained from OECD entrepreneurship and financing statistics. VC provides a useful complementary margin because it is typically influenced strongly by global discount rates and global valuation cycles; thus, it helps distinguish local repricing effects from global financial-cycle co-movement. This aligns with the broader literature on political uncertainty and investment dynamics as well as uncertainty shocks.

### 3.2 Sources and frequency

Our dataset spans the period 2008-2025 at monthly frequency. Financial variables such as 10-year yields and yield spreads are from Bloomberg DataStream and aggregated to monthly frequency. Survey-based indicators and the CLI are from the OECD. For consumer confidence and business tendency indicators, we use monthly growth rates as stated above, consistent with the OECD Consumer Barometer concept for CCI.

### 3.3 Treatment timing and samples

The treatment is the October 7, 2023 shock, treated as a sharply dated extreme political rupture. We set the treatment month to 2023:M10 and study dynamic responses thereafter through 2025M8. The pre-treatment window, ranging from 2008:M1 until 2023:M9, provides a long history to estimate the counterfactual trajectory credibly. The treated unit is Israel. The baseline donor pool comprises the following eleven advanced economies with independent monetary policy regime2, factually unexposed to the terrorist attack on October 7, 2023. This donor pool is intentionally small to reduce the risk of conflating the counterfactual with common shocks as well as the risk of overfitting (Abadie 2021), and is chosen to reflect economies with deep financial market integration and broadly comparable macroeconomic

---

2 Australia, Canada, Denmark, Japan, New Zealand, Norway, South Korea, Sweden, Switzerland, United Kingdom and United States

institutions, while not being directly exposed to the October 7 shock. The goal is to construct a credible counterfactual benchmark for Israel under common global conditions, rather than to benchmark against countries experiencing contemporaneous internal political breakdowns.

Our measurement strategy is closely aligned with the existing work on political risk and extreme-event pricing. First, by emphasizing long-maturity sovereign yields and spreads, we connect to political-uncertainty asset-pricing models and evidence that political events can move discount rates and tail risk premia, as opposed to expected cash flows alone. Second, by using high-frequency survey-based expectations, we follow the modern view that beliefs are primary economic objects and that large political shocks can trigger persistent expectation shifts distinct from realized activity. Third, our interpretation of persistence in market-based risk premia is consistent with the rare-disaster perspective in which changes in perceived tail risk can have first-order asset-pricing consequences. Furthermore, monthly frequency is not a cosmetic feature; it is central to identification. Extreme political ruptures generate sharp belief updates and rapid transitions from disruption to adaptation. Monthly data allow us to (i) locate the timing of the response precisely, (ii) separate immediate disruption from medium-run stabilization, and (iii) test whether market-based repricing persists as an empirical signature of belief revision rather than transitory volatility. Table 1 provides the descriptive statistics for the full treatment sample assessing it against Israel before and after October 7, 2023.

**Table 1**: Full-sample descriptive statistics

|  | Baseline Control Sample | | | Israel | | | |
| --- | --- | --- | --- | --- | --- | --- | --- |
|  | # Of monthly observations | Mean | Std | Before October 7, 2023 | | After October 7, 2023 | |
|  |  |  |  | Mean | Std | Mean | Std |
| Panel A: Outcome variables | | | | | | | |
| 10-year Government Bond Yield | 2,330 | 2.25 | 1.55 | 3.07 | 1.58 | 4.51 | 0.28 |
| Sovereign Risk Premium | 2,330 | -0.428 | 1.251 | 0.583 | 1.175 | 0.248 | 0.378 |
| Consumer Confidence Indicator (CC) | 2,118 | -0.006 | 0.340 | -0.003 | 0.501 | -0.035 | 0.453 |
| Business Tendency Indicator (BT) | 1,271 | 1.242 | 12.890 | 8.159 | 10.518 | 9.261 | 3.723 |
| Composite Leading Indicator (CLI) | 2,371 | 99.96 | 1.35 | 100.22 | 1.87 | 102.03 | 0.55 |

**Figure 1**: Evolution of Sovereign Risk Premium and Forward-Looking Expectations

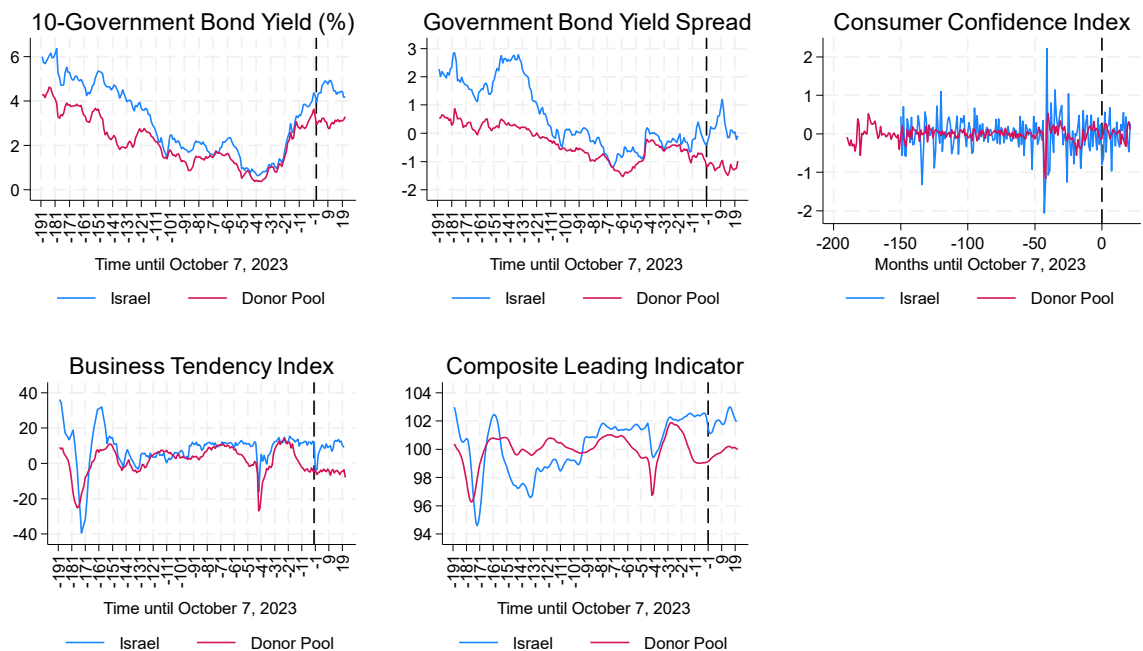

## 4 Identification Strategy and Estimation

### 4.1 The causal problem posed by extreme rupture

The empirical objective of this paper is to identify the causal effects of a sharply dated, rare, and extreme political rupture, the October 7, 2023 shock, on a set of macro-financial and expectation-based outcomes for Israel. This identification problem differs in fundamental ways from those typically studied in the difference-in-differences, event-study, or panel regression literatures ([Freyaldenhoven et. al. 2019](#))

First, the treatment is not randomly assigned. It is unique to a single unit (i.e., Israel) and coincides with a major geopolitical event that plausibly alters beliefs about tail risk, long-run institutional stability, and fiscal sustainability ([Gorodnichenko et. al. 2025](#)). Second, the outcomes of interest, (i.e., sovereign yields, confidence indicators, leading indicators, and venture capital flows), are forward-looking quantities whose dynamics are shaped by expectations, narratives, and belief updating rather than by mechanical adjustment to contemporaneous shocks. Third, the shock occurs in a globally integrated financial system, meaning that global confounders such as monetary tightening cycles, risk-on/risk-off episodes, and valuation cycles, are likely to co-move with the treated unit's outcomes ([Wang et. al. 2024, Choi and Hadad 2025](#)). These features render standard panel regression approaches inadequate. Two-way fixed effects (TWFE) regressions, even when augmented with unit-specific trends, implicitly rely on a strong parallel trends assumption, that Israel's untreated potential outcome path would have followed the same linear (or smoothly parameterized) trend as the donor pool absent the shock. In a setting with large, time-varying global shocks and heterogeneous country loadings, this assumption is implausible. Moreover, TWFE specifications are ill-suited for cases with a single treated unit and dynamic treatment effects ([De Chaisemartin and d'Haultfoeuille 2020](#))

Classic event-study designs encounter similar difficulties. They require that pre-treatment dynamics be well approximated by a parametric form and that no unit-specific deviations emerge absent treatment ([MacKinlay 1997](#)). Yet in macro-financial panels, pre-treatment paths are often shaped by complex, nonlinear interactions between global and domestic factors (Caldara and Iacovello 2022, Caldara et. al. 2025). Forcing these into linear trends risks either overfitting or underfitting, both of which compromise causal inference (Baker and Gelbach 2020). Synthetic control (SC) methods partially address these issues by constructing a weighted combination of donor units that matches the treated unit's pre-treatment path (Abadie 2021). However, standard classical SC imposes strong constraints such that weights must be non-negative, sum to one, and be time-invariant (Botosaru and Ferman 2019, Gilchrist et. al. 2023). These restrictions can be problematic in settings where the treated unit's relationship to global factors evolves over time, as is typical in macro-financial data (Kuruc 2022)

Our strategy builds on a different conceptualization of the problem, treating the counterfactual path of Israel after October 7 as missing data. The econometric task is therefore not to estimate a regression coefficient, but to impute a missing block of a panel matrix under structural assumptions about the untreated outcome process. This perspective naturally leads to the matrix completion framework developed by Athey et al. (2021), which treats causal inference in panels as a structured missing-data problem.

### 4.2 Potential Outcomes, Timing, and Dynamic Estimands

Following Rubin (1983), we formalize the problem using the potential outcomes framework. Let $i \in \{1, \dots N\}$ index countries and $t \in \{1, \dots T\}$ index months. Let $i = 1$ denote Israel and $i \in 2, \dots N$ denote donor countries unexposed to October 7 shock. The shock occurs in month $T_0$ corresponding to October 2023. For each unit and time, we define the potential outcome $Y_{i,t}(0)$ as the outcome that would be observe in the absence of the October 7 shock, and $Y_{i,t}(1)$ as the outcome observed under shock. Then, the observed outcome is modelled as:

$$Y_{i,t} = D_{i,t}Y_{i,t}(1) + (1 - D_{i,t})Y_{i,t}(0)$$

where $D_{1,t} = \mathbf{1}\{t \geq T_0\}$ and $D_{i,t} = 0$ for all $i \geq 2$ and all $t$. The causal quantity of interest is not a single scalar but a full dynamic treatment effect path:

$$\tau_t = Y_{1,t}(1) - Y_{1,t}(0), \qquad t \geq 0$$

This formulation is essential. Because the shock is interpreted as a belief rupture, not merely a transitory disturbance, its effects may evolve over time, exhibiting sharp impact responses, partial reversals, or persistent deviations. Collapsing these dynamics into a single average risk conflating distinct economic mechanisms. Accordingly, we report both dynamic treatment effect $\{\tau_t\}_{t \geq T_0}$ and window-specific average treatment effects denoted as:

$$ATET(\mathcal{T}_1) = \frac{1}{|\mathcal{T}_1|} \sum_{t \in \mathcal{T}_1} \tau_t$$

where $\mathcal{T}_1$ may correspond to short-run, medium-run or long-run windows. The horizon-specific perspective is central to our theoretical framework in Section 2 which emphasizes differential belief updating across time horizons. In turn, the identification problem is immediate and suggests that $Y_{1,t}(0)$ is unobserved for all $t \geq T_0$, and the question that remains is how to recover the counterfactual path in a way that is principled, flexible and disciplined.

### 4.3 The Counterfactual as a Structured Missing-Data Problem

A key conceptual step in this paper is to treat the causal problem as a missing-data problem rather than as a regression problem. Let $Y \in \mathbb{R}^{N \times T}$ denote the matrix of observed outcomes. Define the matrix of untreated potential outcomes $L \in \mathbb{R}^{N \times T}$ where $L_{i,t} = Y_{i,t}(0)$. The observed data consists of all entries of $L$ except those corresponding to Israel in the post-October 7 period. More formally, define the set of observed untreated entries:

$$\Omega = \{(i,t): i \geq 2 \text{ for all } t\} \cup \{(1,t): t \leq T_0\}$$

The set of missing entries is captured by the following Hadamard matrix:

$$\mathcal{M} = \{(1,t): t \geq T_0\}$$

Our task is to impute $\{L_{1,t}\}_{t \geq T_0}$ using the information contained in the observed block $\Omega$. Once the imputation is completed, the treatment effect follows mechanically:

$$\hat{\tau}_t = Y_{1,t} - \hat{L}_{1,t}, \quad t \geq T_0$$

This framing has several crucial implications. First, identification does not come from a parametric functional form, such as linear trends. It comes from structure, in particular, from restrictions on the complexity of the untreated outcome process. Second, the donor pool is not merely a set of "controls." It is the source of information that allows us to learn the latent structure governing the joint evolution of outcomes across countries. Third, the long pre-treatment window is not a convenience; it is a necessity. Without sufficient pre-treatment variation, we cannot learn the latent structure well enough to impute the missing block credibly. This perspective invariably unifies synthetic control properties (Li 2020), interactive fixed effects (Xu 2017), and difference-in-differences under a common missing-data umbrella (Roth e.t al. 2023). What distinguishes these methods is not whether they are "causal," but the type of structure they impose on $L$. Matrix completion imposes that $L$ is low-rank, that is, it can be well approximated by a small number of latent factors with heterogeneous loadings. This assumption is both economically plausible and statistically powerful in macro-financial panels.

### 4.4   Low-Rank Structure

The central identifying restriction underlying matrix completion is that the matrix of untreated potential outcomes, $L$, admits a low-rank approximation. Formally, this means that $L$ can be represented as a finite sum of a small number of latent factors and heterogeneous country-specific loadings, so that $\sum_{k=1}^{r} \lambda_{i,k} f_{k,t}$ for some $r \ll \min(N, T)$. This assumption is not merely technical. It reflects the economic reality that macro-financial outcomes across advanced economies are jointly shaped by a limited set of global forces such as global monetary conditions, risk appetite, geopolitical regimes, and valuation cycles, that evolve over time but load differently across countries.

The low-rank structure is substantially weaker than the parallel-trends assumption underlying conventional difference-in-differences designs. Parallel trends impose that all units share the same untreated time path up to an additive constant or a low-dimensional parametric trend. By contrast, the low-rank restriction allows for nonlinear dynamics, heterogeneous responses to global shocks, and time-varying co-movement patterns. It permits Israel's untreated path to differ flexibly from the donor pool, provided these differences are explained by stable factor exposures learned from the pre-October 7 period.

From a statistical perspective, the low-rank assumption enables recovery of missing entries from observed ones by exploiting shared structure across rows and columns (Chernozhukov et. al. 2023, Choi et. al. 2025). This is the same logic underlying collaborative filtering and recommendation systems. When outcomes are governed by a small number of latent dimensions, unobserved values can be accurately imputed from observed data (Fithian and Mazmuder 2018). In our setting, the long pre-treatment window and the presence of multiple donor units allow the factor structure to be estimated with high precision, making it possible to reconstruct Israel's post-treatment counterfactual path.

### 4.5  Two-Way Fixed Effects and Interactive Heterogeneity

We decompose untreated potential outcomes as:

$$Y_{i,t}(0) = \alpha_i + \gamma_t + L_{i,t} + \varepsilon_{i,t}$$

where $\alpha_i$ denotes country-fixed effects, $\gamma_t$ denotes time-fixed effects and $L_{i,t}$ captures the low-rank interactive component. The unit fixed effects absorb permanent cross-country differences in institutional quality, fiscal credibility, economic structure, and survey response behavior. The time fixed effects absorb global shocks common to all units, such as synchronized monetary tightening, worldwide recessions, or shifts in global risk premia. These additive components are essential: without them, the low-rank structure would be forced to explain systematic level differences and global co-movement, distorting the estimation of the latent factor structure.

The interactive component $L_{i,t}$ captures heterogeneous exposure to common forces and is the core object of interest. It allows different countries to load differently on global risk cycles, geopolitical regimes, and valuation dynamics, and it allows these differences to matter in a time-varying way. This is precisely the form of heterogeneity that dominates macro-financial panels and that cannot be accommodated by two-way fixed effects. Standard two-way fixed effects impose that all units respond identically to global shocks up to a constant. In our setting, this restriction is untenable. Israel's exposure to global factors, geopolitical risk, and security-related uncertainty plausibly differs from that of donor economies, and these differences may evolve over time. By allowing for interactive heterogeneity, the matrix completion framework avoids attributing differential factor exposure to treatment, which is a common failure mode of additive fixed-effects designs.

### 4.6 Matrix Completion Estimator

Following Athey et. al. (2021), we estimate the untreated outcome matrix by solving

$$(\hat{\alpha}, \hat{\gamma}, \hat{L}) = \arg\min_{\alpha, \gamma, L} \sum_{(i,t) \in \Omega} (Y_{i,t} - \alpha_i - \gamma_i - L_{i,t})^2 + \lambda \cdot \|L\|$$

where $\Omega$ denotes the set of observed untreated entries and $\|L\|$ is the nuclear norm of $L$, defined as the sum of its singular values. The nuclear norm acts as a convex relaxation of the rank constraint, shrinking the latent structure toward low dimensionality without imposing a fixed number of factors ex ante. This feature is particularly important in macro-financial panels, where the true number of relevant global forces is unknown and may differ across outcomes. The estimator therefore selects the effective dimensionality of the latent structure endogenously, balancing goodness of fit against model complexity.

This formulation clarifies that matrix completion is not a regression model and does not deliver a "treatment coefficient." Instead, it estimates the joint structure governing untreated outcomes and then interprets post-treatment deviations from this structure as causal effects. The treatment effect is therefore defined residually, as the component of Israel's post-treatment path that cannot be explained by the latent factor structure learned from the donor pool and the pre-treatment period. This perspective is critical. Identification does not arise from conditioning on observables or from parametric extrapolation. It arises from structural regularity in the untreated outcome process. If this structure is stable, the missing post-treatment entries can be imputed credibly. If it is not, no panel estimator, either synthetic control, difference-in-differences, or otherwise, can recover a valid counterfactual.

### 4.7 *Rolling-Window Cross-Validation for the Regularization Parameter*

A central practical and conceptual component of the matrix completion estimator is the choice of the regularization parameter $\lambda$ which governs the complexity of the latent structure. Small values of $\lambda$ allow the low-rank component to absorb a large share of the variation in the data, potentially fitting idiosyncratic noise. Large values impose stronger shrinkage, risking underfitting by over-smoothing genuine heterogeneity. Because treatment effects are defined residually, as deviations from the estimated latent structure, the credibility of the counterfactual path depends critically on how this bias-variance tradeoff is plausibly resolved.

We select $\lambda$ using rolling-window cross-validation on the pre-treatment block. This procedure is designed to mimic the actual imputation problem we face, predicting future untreated outcomes from past information. Specifically, we partition the pre-treatment sample into contiguous training and validation windows, estimate the model on the training window, and evaluate its predictive performance on the subsequent validation window. This procedure is iterated across the pre-treatment sample, and $\lambda$ is chosen to minimize average out-of-sample prediction error.

This approach follows the logic of the matrix completion literature, which emphasizes that the objective is accurate imputation of missing values rather than in-sample fit (Candès and Recht 2009, Athey et al. 2021). It also parallels best practices in time-series cross-validation, where random subsampling is inappropriate due to serial dependence and structural persistence (Bergmeir et. al. 2018). Conceptually, rolling-window validation plays a causal role. Without disciplined regularization, the estimator could inadvertently fit idiosyncratic features of Israel's pre-treatment trajectory that are not shared by the donor pool. Such overfitting would mechanically improve pre-treatment fit but would contaminate the post-treatment counterfactual. Cross-validation ensures that the latent structure is learned from stable cross-unit regularities rather than from unit-specific noise, thereby preserving the interpretability of post-treatment deviations as causal effects.

### 4.8 Identification Assumptions: Formal and Economic Context

The causal interpretation of the imputed counterfactual rests on a small number of structural assumptions. These assumptions formalize the types of regularity that any panel-based counterfactual method must invoke, albeit often implicitly. The first assumption is that untreated outcomes admit a stable low-rank representation. Economically, this means that a limited number of latent global forces jointly shape the evolution of outcomes across advanced economies. This does not require these forces to be constant, only that the dimensionality of the system remain small. This assumption is closely related to factor-structure representations in macroeconomic panels (Bai 2009, Stock and Watson 2016).

The second assumption is stability of factor loadings in the absence of treatment. That is, absent the October 7 shock, Israel's exposure to the latent factors would have continued to evolve in a manner consistent with its pre-treatment relationship to the donor pool. This is the analogue of the parallel-trends assumption, but it is strictly weaker. It allows for non-linear dynamics, heterogeneous responses to global shocks, and time-varying co-movement patterns. The key requirement is not that Israel follow the same path as the donors, but that it continues to reside on the same low-dimensional manifold of untreated outcomes.

The third assumption is no interference. The shock of October 7 in Israel does not directly affect donor outcomes. This is standard in panel causal inference. To the extent that the shock affects global risk sentiment or financial conditions, these common components are absorbed by time fixed effects and latent common factors. The fourth assumption is no anticipation. Outcomes prior to October 2023 are assumed not to embed information about the shock. This assumption is directly testable through pre-treatment diagnostics and placebo exercises.

Finally, the design requires sufficient support in the pre-treatment period to estimate the latent structure. This is where the long pre-period and monthly frequency play a central role. High-frequency data provide repeated realizations of the factor structure under different global conditions, improving the precision with which the mapping from latent factors to observed outcomes is learned. Taken together, these assumptions imply that the untreated potential outcome path of Israel is predictable from the joint distribution of outcomes across the donor pool and Israel's own history. Matrix completion operationalizes this predictability in a flexible, data-driven manner.

*4.9     What Identifies the October 7 Effect?*

It is useful to be explicit about what, exactly, identifies the October 7 effect in our framework. Identification does not arise from a parametric extrapolation, a regression coefficient, or a single comparison. Instead, it arises from a structural inconsistency: after October 2023, Israel's observed trajectory violates the continuation of the low-rank structure learned from the pre-treatment period. The estimator learns a mapping from latent global forces to country-specific outcomes using the entire donor pool and Israel's own pre-treatment history. This mapping implies a specific prediction for how Israel's outcomes should evolve in the absence of treatment, conditional on the realized paths of the latent factors. If Israel's post-shock outcomes systematically deviate from this prediction, the only remaining explanation, under the maintained assumptions, is the treatment itself.

In this sense, the treatment effect is not imposed by construction; it emerges as a residual. It is the component of Israel's post-treatment path that cannot be explained by any low-dimensional combination of global forces consistent with its pre-treatment behavior. This logic is closely related to the identification arguments underlying interactive fixed effects models (Bai 2009) and synthetic control methods (Abadie et. al. 2010), but with substantially greater flexibility. This perspective is particularly well suited to belief-driven shocks. Belief ruptures do not merely shift levels; they alter how agents interpret and price future states. Such changes manifest precisely as violations of previously stable co-movement patterns, not necessarily as sharp one-time jumps. By allowing both the factors and the loadings to evolve flexibly, matrix completion distinguishes between genuine structural breaks and ordinary differential exposure to global shocks. In short, the October 7 effect is identified as the failure of Israel's post-treatment path to remain on the low-dimensional manifold of untreated outcomes.

*4.10    Dynamic Treatment Effects and Horizon Decomposition*

Because the October 7 shock is interpreted as a rupture in beliefs rather than a transitory disturbance, it is essential to allow treatment effects to evolve over time. We therefore do not summarize the impact of the shock with a single scalar coefficient. Instead, we estimate the full dynamic treatment effect path $\{\tau_t\}_{t\geq 0}$, where $\tau_t = Y_{1,t}(1) - Y_{1,t}(0)$

This event-time representation allows us to distinguish between several economically distinct components of the response. Immediate movements capture short-run disruption, panic, and liquidity effects. Medium-run movements reflect adaptation, learning, and policy responses. Persistent deviations correspond to durable belief revision and long-horizon repricing. To formalize this distinction, we report horizon-specific averages. Let $\mathcal{T}^{(S)}$, $\mathcal{T}^{(M)}$ and $\mathcal{T}^{(L)}$ denote short-, medium-, and long-run post-treatment windows. We compute:

$$ATET^{(h)} = \frac{1}{|\mathcal{T}^{(h)}|} \sum_{t \in \mathcal{T}^{(h)}} \tau_t, \qquad h \in \{S, M, L\}$$

This decomposition is not cosmetic. It is vital for interpretation. Many political shocks generate large immediate reactions that dissipate, while others permanently alter the perceived distribution of future states. By allowing for horizon-specific effects, we avoid conflating these mechanisms. This perspective aligns with recent work emphasizing that macro-financial responses to political events often involve persistent belief revision rather than transitory volatility (Pastor and Veronesi 2013, Gennaioli et. al. 2018). It also parallels the rare-disaster literature, where small changes in perceived tail probabilities can produce large and persistent asset price movements even if near-term fundamentals recover (Barro 2006, Gabaix 2012).

### 4.11 Pre-Shock Diagnostics and Internal Validity

Because matrix completion is fundamentally an imputation-based design, pre-treatment fit is not merely a descriptive diagnostic; it is part of the identifying logic. The estimator learns the latent structure of untreated outcomes from the pre-treatment block, and the credibility of post-treatment imputation depends on how well this structure reproduces Israel's pre-treatment path.

Let $\hat{Y}_{i,t}(0)$ denote the imputed counterfactual for Israel in the pre-October 7 period. Define the pre-treatment residuals:

$$\delta_t = Y_{1,t}(1) - Y_{1,t}(0), \quad t < T_0$$

Under correct specification, these residuals should be mean-zero and exhibit no systematic temporal pattern. We assess pre-treatment fit using three complementary tools. First, we present visual comparisons of observed and imputed paths. Visual diagnostics are particularly informative in dynamic settings, where misspecification often manifests as systematic curvature or phase shifts rather than isolated outliers. Second, we report quantitative measures of fit such as the mean squared prediction error (MSE) over the pre-treatment window. This provides a summary measure of how well the latent structure approximates Israel's untreated path. Third, we conduct formal tests of the null hypothesis that pre-treatment residuals are jointly zero. Rejection of this null indicates that the model is systematically mis-predicting Israel even before treatment, which undermines the credibility of post-shock imputation.

Importantly, we do not interpret a rejection of this null mechanically as a failure of the design. In macro-financial panels, perfect pre-fit is neither attainable nor desirable. Overfitting the pre-period can artificially suppress post-treatment gaps. What matters is that pre-treatment deviations be small, unsystematic, and economically negligible relative to post-treatment movements. This emphasis on pre-treatment diagnostics parallels the logic of synthetic control methods, where credibility hinges on how well the synthetic unit reproduces the treated unit prior to treatment (Abadie et. al. 2010). In the matrix completion framework, however, this logic is formalized within a flexible, high-dimensional structure rather than through fixed convex weights.

### 4.12 Inference: In-Space and In-Time Placebos

Standard asymptotic inference is inappropriate in our setting. The panel contains a single treated unit, treatment is sharply dated, and outcomes exhibit serial dependence. Conventional heteroskedasticity-robust or clustered standard errors do not capture the relevant sources of uncertainty. Instead, we rely on design-based inference, following the logic of permutation and placebo tests developed in the synthetic control and panel causal inference literatures. Our primary inference tools are in-space and in-time placebos.

In-space placebos assign the treatment to each donor country in turn, re-estimating the matrix completion model and computing placebo treatment effects. This generates a distribution of effects under the null of no treatment. Israel's estimated effect is then evaluated relative to this distribution. If Israel's effect lies in the extreme tail, we reject the null that the observed deviation could plausibly arise from latent factor variation alone. In-time placebos assign pseudo-treatment dates within the pre-treatment period for Israel and recompute treatment effects. This tests whether the estimator systematically generates spurious "effects" even in periods without true treatment. A credible design should produce near-zero placebo effects in these exercises.

Together, these two forms of placebo inference address different threats. In-space placebos test whether Israel is unusually extreme relative to donor units. In-time placebos test whether the design falsely detects treatment effects when none exist. This approach follows the design-based logic emphasized by Abadie et al. (2010), Firpo and Possebom (2018), and Arkhangelsky et al. (2021), among others, who argue that in settings with few treated units, inference should be grounded in permutation-style reasoning rather than large-sample approximations. We report p-values computed as the fraction of placebo effects that exceed the magnitude of Israel's estimated effect. This procedure yields a transparent, finite-sample measure of statistical significance that does not rely on parametric assumptions.

### 4.13   Relationship to Alternative Estimators

It is useful to clarify how matrix completion relates to other commonly used panel estimators and why it is particularly well suited to our setting. Traditional difference-in-differences estimators rely on a parallel-trends assumption. In the absence of treatment, the treated unit would have followed the same trend as the control group. In macro-financial panels with heterogeneous exposure to global shocks, this assumption is implausible. Countries differ not only in levels but in how they load on global forces, and these loadings may evolve over time. Two-way fixed effects models impose this restriction implicitly, and their failure in such environments is well documented. Synthetic control methods relax parallel trends by constructing a weighted average of control units that matches the treated unit's pre-treatment path. However, standard synthetic control constrains weights to be nonnegative, time-invariant, and summing to one. These constraints limit flexibility and can lead to poor performance when the treated unit's relationship to global factors changes over time or when no convex combination of donors reproduces its dynamics.

Interactive fixed effects models (Bai 2009, Xu 2017) allow for latent factors and heterogeneous loadings, making them conceptually close to matrix completion. The difference lies in implementation. Interactive fixed effects typically require the researcher to choose the number of factors ex ante and estimate them via principal components. Matrix completion, by contrast, uses nuclear-norm regularization to let the data determine the effective dimensionality of the latent structure. This is particularly valuable in settings where the number of relevant latent dimensions is unknown and may differ across outcomes. Against this backdrop, synthetic difference-in-differences (SDID) estimator combines features of synthetic control and difference-in-differences by constructing unit weights and time weights that balance treated and control units along both dimensions (Arkhangelsky et al. 2021). We use SDID as a robustness check. Its identifying logic differs from that of matrix completion, relying more heavily on weighted averaging and differencing rather than on low-rank imputation.

Agreement between the two approaches therefore provides strong evidence that our results are not an artifact of a particular estimator.

From a unifying perspective, many modern panel estimators can be seen as special cases of regularized matrix approximation under different constraints. Matrix completion occupies a particularly flexible point in this space by allowing for time-varying heterogeneity, nonlinear dynamics, and complex comovement patterns without imposing rigid parametric structure. These features are essential in a setting where outcomes are driven by belief updating rather than by mechanical adjustment to contemporaneous shocks.

### 4.14  *Threats to identification and prospective mitigation*

No observational design is free of threats to identification, and it is important to be explicit about where ours could fail. One concern is donor contamination. If the October 7 shock affected donor countries directly, through trade channels, geopolitical exposure, or security spillovers, then the donor pool would no longer represent untreated outcomes. We mitigate this by selecting donor countries with minimal direct exposure and by absorbing global shocks through time fixed effects and latent factors. We further assess robustness to alternative donor subsets.

A second concern is global financial contagion. If the shock triggered a worldwide repricing of geopolitical risk, then some of the effect would be absorbed by common components rather than attributed to Israel. This would bias our estimates downward rather than upward. Our interpretation of the results therefore emphasizes that estimated effects should be understood as Israel-specific deviations from global repricing. A third concern is structural instability of the latent factor structure. If the dimensionality or nature of global forces changed sharply at the time of treatment, imputation could fail. While this cannot be ruled out entirely, the stability of pre-treatment fit, the consistency across outcomes, and the placebo diagnostics all speak against this possibility.

A fourth concern is endogenous anticipation. If agents anticipated the shock well in advance, pre-treatment outcomes might already embed some of its effects. We address this through in-time placebo tests and by examining whether pre-treatment residuals display systematic drift. Finally, there is the risk of overfitting. Because treatment effects are defined residually, any overfitting of the pre-treatment block mechanically shrinks post-treatment gaps. Our use of rolling-window cross-validation and nuclear-norm regularization is designed to guard against this. These threats are not unique to matrix completion. They are endemic to all panel-based counterfactual designs. What distinguishes our approach is that each of these threats has a transparent diagnostic of pre-treatment fit, placebo distributions, donor sensitivity, and horizon-specific effects.

*4.15    Why this Design is Well Suited to Belief Shocks*

The identification strategy is closely aligned with the conceptual framework developed in Section 2. That framework emphasizes that extreme political ruptures operate primarily through belief updating rather than through immediate changes in productive capacity. Beliefs about tail risk, institutional fragility, and long-run stability alter how agents price assets, form expectations, and allocate capital. Belief shocks are not well captured by linear trends or static comparisons. They often manifest as gradual divergence from previously stable co-movement patterns, rather than as discrete jumps. Matrix completion is particularly well suited to detecting such changes because it learns the joint structure of untreated outcomes and flags deviations from that structure as causal.

This is especially important for long-horizon outcomes such as sovereign yields, which respond to revisions in perceived future states rather than to contemporaneous fundamentals. It is equally important for survey-based expectations, which can remain depressed even when real activity recovers. By allowing for flexible latent dynamics, the estimator distinguishes between global shifts in sentiment and country-specific belief ruptures. The use of monthly data further strengthens this alignment. Belief updating is continuous, not quarterly. High-frequency data allow us to observe the speed of adjustment, the presence of overshooting, and the persistence of deviations. Such patterns are invisible at lower frequencies. In this sense, the identification strategy is not merely a choice but a straightforward and plausible extension of the theory.

*4.16    Summary and Overview*

This section has described a flexible and disciplined approach to identifying the causal effects of the October 7 shock on Israeli macro-financial outcomes. By framing the problem as one of structured missing data, we leverage the low-rank nature of untreated outcomes to impute credible counterfactual paths. Unit and time fixed effects absorb permanent heterogeneity and global shocks, while nuclear-norm regularization allows the latent structure to be learned from the data. Identification arises from violations of previously stable co-movement patterns, not from parametric extrapolation. Rolling-window cross-validation guards against overfitting. Pre-treatment diagnostics ensure internal validity. Placebo-based inference provides finite-sample credibility. Taken together, these features make the design particularly well suited to the study of belief-driven political shocks, precisely the key object of interest in this paper.

## 5    Results

*5.1    Overview: A Belief-Rupture Pattern Across Horizons*

The October 7 shock generates a striking and internally coherent reordering of expectations across horizons. Our results suggest that in sovereign debt markets, Israel experiences a large increase in long-maturity yields and in the sovereign risk premium relative to the United States, indicating a repricing of long-run tail risk rather than a purely global interest-rate movement. In parallel, households' forward-looking beliefs deteriorate sharply, as captured by a statistically tight decline in consumer confidence growth. Yet at the same time, firm-facing and medium-horizon indicators move in the opposite direction. Business tendency measures and the OECD composite leading indicator (CLI) rise strongly, and venture capital flows do not collapse relative to the donor pool.

This cross-outcome pattern is not an empirical curiosity. It is precisely what the conceptual framework in Section 2 predicts when an extreme rupture primarily operates through belief updating about the distribution of future states. Under such a shock, the long-horizon "state-contingent" outcome of interest, principally sovereign risk premia and household welfare beliefs, should respond differently from medium-horizon momentum measures (CLI) and short-horizon operational expectations, captured by business tendency dynamics. The estimated treatment-effect paths provide direct evidence of this horizon separation. In the pre-October 7 period, estimated gaps are close to zero and show no systematic drift, the discontinuity emerges sharply at the October 7 cutoff, and the subsequent dynamics differ by outcome in a way consistent with the belief-horizon mechanism rather than with a single-factor macro shock. The remainder of Section 5 begins with the sovereign-risk results because they serve as the paper's cleanest market-based sufficient statistic for tail-risk repricing. We then interpret the remaining outcomes in later subsections as a structured propagation of the shock across belief horizons.

### 5.2 Long-Horizon Repricing: 10-Year Yields and the Sovereign Risk Premium

#### 5.2.1 Baseline magnitudes and strength

Table 2 reports the baseline matrix-completion estimates for Israel's 10-year government bond yield and the sovereign risk premium (the 10-year yield spread relative to the United States) over the monthly window M1:2008 through M8:2025. The estimated post-October 7 increase in the 10-year yield is economically large and statistically significant at conventional levels. Under low-shrinkage parameter regime, the estimated average effect is approximately 0.56-0.59 percentage points, respectively. Under higher shrinkage, it rises to roughly 0.79 percentage points. The corresponding risk premium increases are of similar magnitude of about 0.60 percentage points under low-shrinkage regime and about 0.79 percentage points under high-shrinkage counterpart. These estimates are stable as the number of bootstrap replications runs increases from 100 to 10,000, and the simulation-based p-values are uniformly small, often effectively very near zero and uniformly within 5% significance threshold.

Two points matter for interpretation. First, the close similarity between yield effects and spread effects implies that the repricing is not merely a reflection of global rate movements. The spread isolates the country-specific premium component, and it moves sharply as well. Second, the sensitivity to the regularization regime is informative rather than alarming. Higher shrinkage imposes a more parsimonious latent structure and mechanically, it attributes less of Israel's post-October 7 variation to latent factors and more to the treatment. We therefore interpret the low- versus high-shrinkage estimates as a disciplined sensitivity band and a range of plausible effect sizes under more or less aggressive factor shrinkage, rather than as competing point truths.

**Table 2**: Effect on October 7 Attack on 10-Year Government Bond Yield and Risk Premium, M1:2008-M8:2025

|  | 10-Year Government Bond Yield | | | Sovereign Risk Premium (Bond Yield Spread) | | |
|---|---|---|---|---|---|---|
|  | (1) | (2) | (3) | (4) | (5) | (6) |
| # bootstrap runs | 100 | 1,000 | 10,000 | 100 | 1,000 | 10,000 |
| Panel A: Low hyper-space shrinkage parameter ($\lambda = 0.0001$) | | | | | | |
| $\widehat{\mu_{T_0+k}}$ | .586*** | .563*** | .563*** | .604*** | .604*** | .604*** |
|  | (.196) | (.195) | (.190) | (.164) | (.170) | (.177) |
| Simulation-based p-value | 0.004 | 0.004 | 0.003 | 0.000 | 0.000 | 0.001 |
| Panel B: High hyper-space shrinkage parameter ($\lambda = 0.001$) | | | | | | |
| $\widehat{\mu_{T_0+k}}$ | .786*** | .786*** | .787*** | .791*** | .791*** | .791*** |
|  | (.173) | (.166) | (.154) | (.166) | (.157) | (.154) |
| Simulation-based p-value | 0.000 | 0.000 | 0.000 | 0.000 | 0.000 | 0.000 |

*Notes*: This table reports baseline matrix completion estimates of the effect of the October 7, 2023 shock on Israel's 10-year government bond yield and sovereign risk premium. The post-rupture outcomes for Israel are treated as missing data and their counterfactual paths are imputed using a low-rank latent factor structure estimated from the pre-treatment period and the donor pool. Panel A and Panel B report estimates obtained under alternative values of the hyper-space shrinkage parameter, illustrating the sensitivity of the results to the degree of regularization. Across specifications, the estimator includes unit and time fixed effects and employs nuclear-norm regularization. Columns differ by the number of bootstrap or simulation runs used to compute inference. Inference is conducted using simulation-based placebo procedures. Reported p-values on null hypothesis are empirical and finite-sample valid. Asterisks denote statistical significance at the 10% (*), 5% (), and 1% (*) levels.

### 5.2.2 Dynamic responses: persistence, hump-shapes, and partial reversal

The dynamic estimates, depicted in Figure 1, reveal that the sovereign-risk response is not a one-month blip. For the 10-year yield, the estimated effect rises immediately after treatment and continues to build over the subsequent months, suggesting that the repricing was not fully resolved by an instantaneous "news update," but rather evolved as beliefs and assessments of long-run states were revised. For the sovereign risk premium, the dynamic pattern is especially revealing. The spread response increases sharply after October 7, reaches a pronounced peak roughly around the first year following the shock, and then partially declines, remaining elevated relative to the pre-rupture period. This hump-shaped premium is exactly what one expects when the market initially reprices tail risk aggressively under severe uncertainty, and then partially retraces as new information arrives and uncertainty is resolved, without returning to the pre-shock baseline. Economically, this is consistent with a belief shock that changes perceived downside probabilities in a persistent way while allowing for some learning-driven mean reversion as the post-event environment becomes better understood.

This dynamic structure matters for our central claim. A pure liquidity shock would tend to generate a sharp spike and rapid reversal. A purely transitory risk-off episode would likely move the level of yields but not sustain a relative premium against the U.S. benchmark. What we observe instead is a sustained re-ranking of Israel's long-horizon pricing relative to comparable advanced economies, with the spread dynamics indicating both an immediate reassessment and a slower process of belief stabilization.

**Figure 1**: Effect of October 7 Attack on Government Bond Yield and Risk Premium, M1:2008-M8:2025

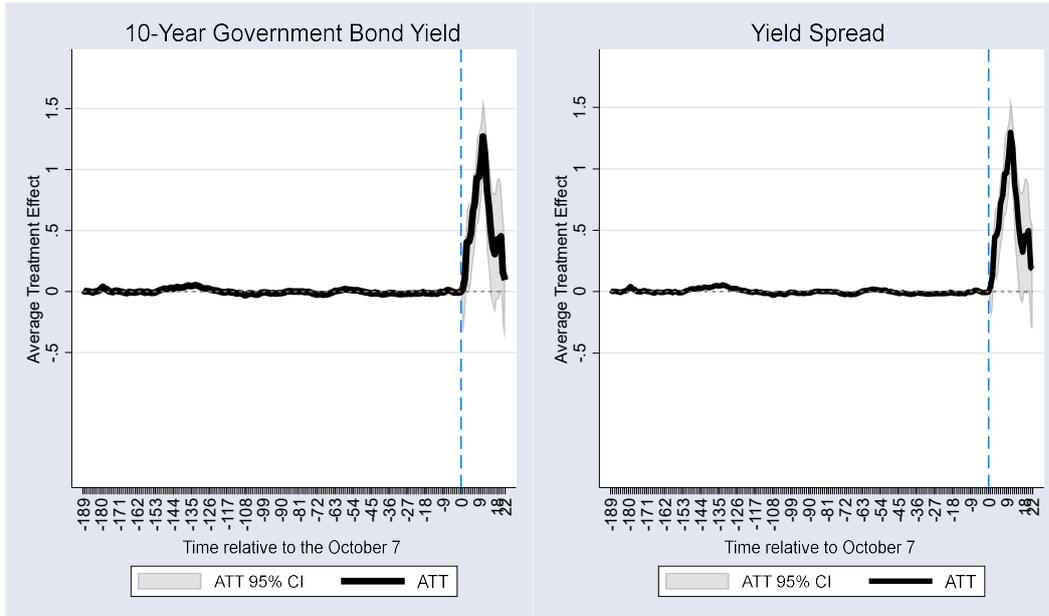

### 5.2.3 Economic interpretation: sovereign yields as a tail risk-sufficient statistic

Long-maturity sovereign yields embed beliefs about future fiscal capacity, political stability, and extreme downside states. In the model developed in Section 2, the key mechanism is that extreme political ruptures increase perceived probability mass in adverse future states, raising the required compensation for holding long-duration domestic sovereign claims. The estimated increase in Israel's 10-year yield and spread is therefore naturally interpreted as a rise in the *sovereign risk premium*, a market-based sufficient statistic for perceived long-horizon tail risk. It is useful to translate magnitudes into economic significance without overstating precision. An increase between 60 and 80 basis points (*bps*) in the sovereign risk premium is material for a government that continuously refinances and rolls over debt along the maturity spectrum. As a rough scaling, a 70 bps increase applied to $100 billion of debt being refinanced at the margin corresponds to about $0.7 billion in additional annual interest cost on that marginal refinancing flow whilst the exact fiscal incidence depends on maturity structure, rollover speed, and hedging. The key point is not the accounting identity but the fact that estimated repricing is large enough to plausibly shift fiscal space and the pricing of domestic long-duration risk for years, not weeks.

### 5.2.4 Internal validity: placebo-based evidence and absence of spurious gaps

The credibility of the sovereign-risk results is reinforced by the in-time placebo tests in Table 3. When pseudo-treatment dates are assigned within the pre-treatment period, estimated "effects" on the 10-year yield and the spread are negative and statistically indistinguishable from zero, with empirical confidence intervals comfortably spanning zero and simulation-based p-values far from rejection. This evidence is important because it speaks directly to the key threat in imputation-based designs, namely, that the estimator might generate spurious post-rupture gaps even absent a genuine structural break. Instead, the placebo results indicate that the model reproduces Israel's pre-treatment relationship to the donor pool sufficiently well that pseudo-shocks do not mechanically produce large quasi-effects.

**Table 3**: Falsifying the timing of October 7 rupture and in-time placebo analysis

|  | 10-Year Government Bond Yield | | Sovereign Risk Premium (Bond Yield Spread) | |
|---|---|---|---|---|
|  | (1) | (2) | (5) | (6) |
| # bootstrap runs | Small-scale (#100) | Large-scale (#1,000) | Small-scale (#100) | Large-scale (#1,000) |
| Panel A: Structural matrix completion setup ($\lambda = 0.0001$) | | | | |
| $\widehat{\mu_{T_0+k}}^{Placebo}$ | -.233 | -.233 | -.209 | -.209 |
|  | (.162) | (.184) | (.165) | (.170) |
| 95% Empirical confidence intervals | (-.606, .102) | (-.607, .105) | (-.554, .057) | (-.556, .091) |
| Simulation-based p-value | {0.246} | {0.199} | {0.188} | {0.218} |

Notes: This table reports in-time placebo estimates designed to falsify the timing of the October 7, 2023 rupture. Pseudo-treatment dates are assigned within the pre-treatment period, and the matrix completion estimator is re-estimated using the same specification, donor pool, and regularization parameters as in the baseline analysis. The reported coefficients represent average post-placebo deviations of Israel's outcomes from their imputed counterfactual paths under false treatment timing. Confidence intervals and p-values on null hypothesis are obtained using simulation-based inference based on repeated placebo assignments. The absence of large or persistent placebo effects provides evidence against spurious detection of breaks and supports the validity of the baseline identification strategy. Asterisks denote statistical significance at the 10% (*), 5% (), and 1% (*) levels.

*5.2.5 Cross-estimator corroboration: synthetic difference-in-differences as a robustness check*

Finally, the sovereign-risk findings are not unique to matrix completion. Table 5 reports synthetic difference-in-differences (SDID) estimates that are closely aligned in magnitude: roughly 0.79 percentage points for the 10-year yield and about 0.77 percentage points for the spread, depending on bootstrap scale. Although SDID confidence intervals are wider and p-values more conservative, the point estimates corroborate the main conclusion that the October 7 shock led to a sizable increase in Israel's long-horizon cost of sovereign capital relative to the donor pool. The SDID weights place most mass on a small subset of donor countries, particularly Australia, Denmark, South Korea, and Sweden, indicating that the counterfactual is being constructed from a coherent "advanced, safe-asset-like" comparison set rather than less pronounced and more diffuse averaging. Taken together, the sovereign-yield and spread results provide the paper's cleanest market-based evidence that October 7 induced a substantial repricing of Israel's long-horizon tail risk. The magnitude is economically meaningful, the dynamics are consistent with belief revision and partial learning, the pre-treatment diagnostics do not suggest spurious drift, and an alternative modern estimator yields similar point estimates.

*5.3   Household Welfare Beliefs: Consumer Confidence and the Persistence of Subjective Risk*

A central implication of the conceptual framework in Section 2 is that extreme political ruptures should operate not only through market pricing of tail risk, but also through household belief updating about future welfare states. Consumer confidence is therefore not merely an auxiliary outcome. It is a direct empirical window into the welfare-belief channel in which households revise expectations about job security, purchasing power, taxation, inflation, public safety, and the general stability of future living standards. Unlike sovereign yields, which aggregate pricing by marginal investors, consumer confidence captures the distribution of beliefs at the household level, and thus provides an important complement to the sovereign risk premium.

*5.3.1  Baseline magnitudes and precision*

Table 4 reports the results of October 7 shock on consumer confidence and other expectation-based indicators. The matrix completion estimates indicate a sharp and statistically tight decline in consumer confidence growth following October 7. The estimated average effect is approximately -0.053 with a narrow confidence interval and effectively zero p-value. This is an unusually precise result for an expectations-based macro indicator in a single-treated-unit setting, and it is consistent with the visual evidence: the post-treatment gap opens immediately at the event date and remains substantially negative thereafter. A key point for interpretation is that the consumer confidence result is not small simply because it is reported in growth rates. Confidence indices are typically scaled and smoothed series, and month-to-month growth rates are designed to capture shifts in sentiment rather than levels. A persistent negative growth-rate effect therefore implies a sustained deterioration in the level of confidence relative to the counterfactual path.

**Table 4**: Effects of October 7 shock on consumer confidence, business tendency and venture capital investments

|  | Consumer confidence dynamics (growth rate) | Business tendency dynamics (growth rate) | Venture capital investment | Composite leading indicator |
|---|---|---|---|---|
|  | monthly | monthly | annual | monthly |
|  | (1) | (2) | (3) | (4) |
| Panel A: Low hyper-space shrinkage parameter ($\lambda=0.0001$) | | | | |
| $\mu_{T_0+k}$ | -.061*** | 7.635*** | .296** | 1.949*** |
|  | (.013) | (2.914) | (.116) | (.149) |
| 95% Empirical confidence intervals | (-.091, -.041) | (1.983, 12.685) | (.066, .489) | (1.603, 2.198) |
| Simulation-based p-value | 0.000 | 0.009 | 0.011 | 0.000 |

Notes: This table reports matrix completion estimates of the effect of the October 7, 2023 shock on consumer confidence dynamics, business tendency dynamics, venture capital investment, and the composite leading indicator. The post-rupture outcomes for Israel are treated as missing data and their counterfactual paths are

*5.3.2 Dynamic response: stickiness and the absence of rapid mean reversion*

The dynamic treatment effect path is notable for its persistence. The post-treatment period is characterized by large volatility in the estimated monthly effects, but the mean effect remains negative throughout, and there is no clear indication of rapid mean reversion back toward the pre-treatment baseline. This is theoretically informative. A purely contemporaneous disruption, say, a temporary supply shock or a short-lived liquidity event, would be expected to generate a sharp drop followed by rebound. The evidence instead suggests a more durable shift in household beliefs. This persistence aligns with the belief mechanism formalized in Section 2. Households revise the perceived probability of adverse future states and update welfare-relevant beliefs slowly, because belief updating is constrained by limited information, narrative framing, and salience. Even if objective conditions partially stabilize, households may continue to assign higher probability to adverse states for a prolonged period, producing sticky confidence effects.

*5.3.3 Connecting household confidence to the sovereign premium: two lenses on the same belief shock*

The consumer confidence findings also help discipline the interpretation of the sovereign yield and spread effects. The sovereign risk premium reflects the pricing of long-horizon tail risk by marginal investors; consumer confidence reflects the perceived welfare risk by households. In a standard macro shock, one might expect these to move together in sign but differ in magnitude. In a belief rupture, however, the combination of an elevated sovereign premium and persistently depressed household confidence is especially diagnostic. It indicates that both the pricing kernel of long-horizon capital markets and the subjective welfare beliefs of households shift in the direction of higher perceived downside risk. Importantly, the consumer confidence result also speaks to a natural alternative explanation for the sovereign spread increase, namely, that the spread reflects purely technical or market microstructure effects. If that were the case, it would be difficult to rationalize why households' welfare beliefs shift so strongly and persistently at the same event date. The household evidence therefore provides cross-outcome corroboration that October 7 induced a deep repricing of perceived future states rather than a narrow market dislocation.

### *5.3.4  Economic meaning: welfare beliefs, precautionary behavior, and feedback effects*

The household channel matters not only as a measure of "sentiment," but because it plausibly shapes real behavior through precautionary saving, reduced discretionary consumption, and labor supply adjustments. In the conceptual framework, heightened perceived downside risk increases precautionary motives and reduces the marginal propensity to consume out of current income. Even if short-run fiscal mobilization supports aggregate demand, precautionary behavior can dampen private demand and amplify macro volatility. This points to a critical contribution of the paper. It identifies a belief shock that simultaneously raises long-horizon risk premia and deteriorates household welfare beliefs, creating the conditions for endogenous feedback between risk pricing and real behavior. This feedback is precisely what standard single-outcome studies miss. It also clarifies why the sovereign premium response should not be interpreted narrowly as "financial market overreaction." The persistence in household beliefs implies that the repricing reflects a broad reweighting of future states across the economy.

### *5.4  Firms, Mobilization, and Adaptation: Business Tendency as a Short-Horizon Mechanism*

We next turn to firm-facing expectations. The business tendency indicator, measured in growth rates, captures near-term business conditions and expectations about production, orders, and demand. This outcome is explicitly short-horizon: it reflects what firms believe will happen over the coming months rather than over the coming years. As argued in Section 2, extreme political ruptures can generate an initial panic response but then trigger rapid adaptation mechanisms, including mobilization-related procurement, fiscal expansion, substitution toward domestic suppliers, and reallocation of economic activity. The business tendency results provide evidence consistent with this adaptation channel.

*5.4.1 Baseline estimates and aggregate uncertainty*

The estimated effect on business tendency is positive, but the inference is notably weaker than for sovereign yields or consumer confidence. The point estimate is approximately +1.50 but the confidence interval is wide and the p-value does not reject the null at conventional levels. This implies that the business tendency series is noisier and that the donor-based imputation is less precise, which is unsurprising for high-frequency firm sentiment measures. Nevertheless, the sign and the time-profile are informative when interpreted jointly with the other outcomes. The key insight is not that firms become unequivocally "optimistic," but that the short-horizon business-expectations margin does not collapse in the way one might expect if the shock were operating purely through a standard contractionary demand channel.

*5.4.2 Reconciling firm-side resilience with household pessimism and sovereign repricing*

At first glance, a positive or non-negative firm tendency response may seem to conflict with the negative consumer confidence effect and the rise in the sovereign premium. In fact, the divergence is exactly what the belief-horizon framework predicts. Households and sovereign investors focus on long-horizon welfare and tail risk. Firms, in contrast, may anticipate substantial near-term demand and activity arising from mobilization and policy responses.

Government procurement, defense-related activity, reconstruction spending, and substitution away from disrupted supply chains can create near-term demand even as long-horizon risk premia rise. In this sense, the business tendency outcome is a short-horizon "adaptation" statistic, while sovereign spreads are long-horizon "tail-risk" statistics, and consumer confidence is a household welfare-belief statistic.

This decomposition is important for two reasons. First, it helps rule out competing explanations. A pure demand contraction should depress firm expectations and household confidence simultaneously. A pure risk-off episode should move spreads and perhaps confidence, but would not naturally generate medium-horizon momentum. The fact that short- and medium-horizon indicators are not uniformly negative strengthens the interpretation that the economy experiences risk-growth decoupling following a belief rupture. Second, it shows why the paper's multi-outcome design is not a collection of robustness checks but an essential part of identification. The pattern across outcomes is itself the evidence: it reveals which horizons of belief move and which do not.

*5.4.3 Implications for the propagation mechanism*

Interpreting these results through the model implies a specific propagation mechanism. Immediately after the shock, long-horizon tail risk increases and is priced into sovereign spreads. Households update welfare beliefs downward and remain pessimistic. Firms, however, anticipate or experience an increase in near-term activity due to mobilization and adjustment, preventing a collapse of short-horizon expectations. Over time, the coexistence of elevated tail risk and sustained near-term activity implies that the economy may exhibit positive momentum in selected indicators even as risk premia remain elevated. This mechanism is consistent with the broader empirical finding that the sovereign premium response appears persistent but not purely monotone: markets reprice sharply, then partially retrace as learning occurs, while household beliefs remain depressed. Figure 2 presents the estimated impacts of October 7 on the full set of outcomes.

**Figure 2**: Matrix completion estimates of the effect of October 7 on consumer and business confidence

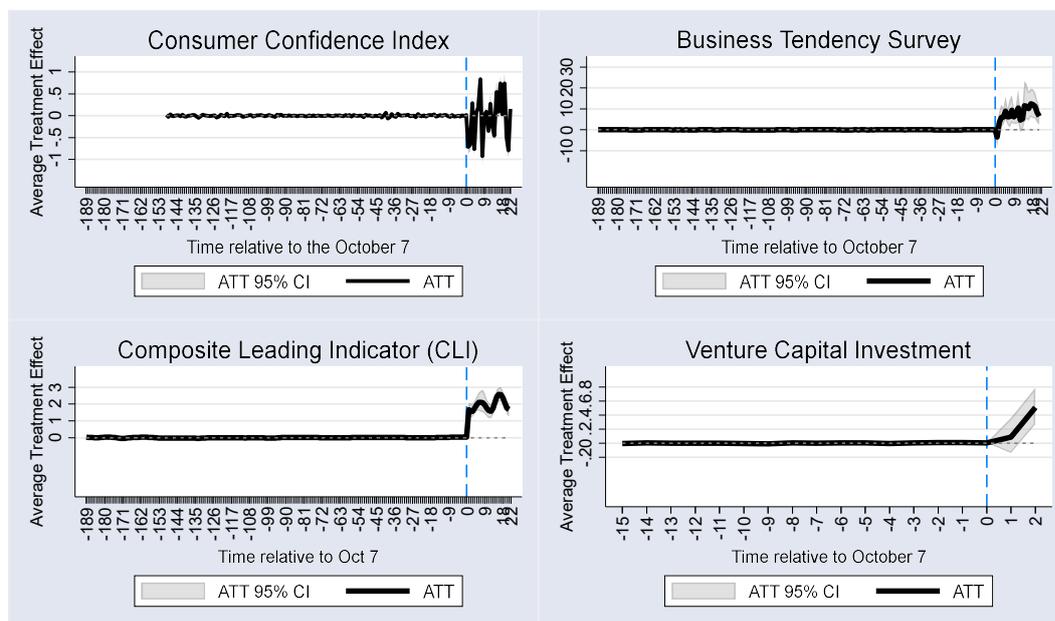

### 5.5 Medium-Run Momentum and Risk-Growth Decoupling: The Composite Leading Indicator

The composite leading indicator (CLI) provides a particularly revealing bridge between the sovereign-risk results and the household and firm belief channels. Whereas consumer confidence reflects households' welfare beliefs and sovereign spreads reflect long-horizon tail-risk pricing, the CLI is designed to anticipate medium-run cyclical momentum. It is therefore neither a pure "sentiment" measure nor a direct measure of tail risk. It is a forward-looking signal about real activity over an intermediate horizon. This distinction is essential for interpreting the results. The conceptual framework in Section 2 predicts that after an extreme political rupture the economy may exhibit risk-growth decoupling where long-horizon risk premia can rise because tail risk and uncertainty increase, even while medium-run activity expectations improve due to mobilization, reallocation, and policy response. The CLI results provide direct empirical support for this mechanism.

#### 5.5.1 Baseline magnitude and sharpness of the discontinuity

The estimated effect of October 7 on the CLI is strongly positive and precisely estimated. The matrix completion estimates indicate an average post-treatment increase of approximately +1.95 in the CLI dynamics, with a narrow confidence interval. The dynamic path shows an immediate discrete jump at treatment, followed by sustained elevation throughout the post-treatment window. Two aspects of this result are noteworthy. First, the effect is large relative to pre-treatment fluctuations, which are tightly centered near zero. Second, the response is not a transient one-month anomaly. The CLI shifts to a higher level of medium-run momentum and remains there, which suggests a systematic reweighting of medium-horizon growth expectations rather than noise.

### 5.5.2 Why a positive CLI effect is not counterintuitive?

At first glance, a positive medium-run leading indicator after an extreme shock may appear counterintuitive. The apparent tension dissolves once we recognize that the CLI is not a measure of welfare, nor is it a measure of tail risk. It is a measure of prospective momentum. In environments characterized by rapid mobilization and substantial reallocation of economic activity, momentum can rise even when perceived long-run risk increases. The October 7 shock plausibly triggers multiple mechanisms that raise medium-run momentum. Mobilization and defense spending can stimulate near-term production and employment. Supply-chain disruption can induce substitution toward domestic production. Heightened demand for certain services and technologies, including security and cyber resilience, can create sector-specific booms that propagate into aggregate indicators. Finally, fiscal and monetary policy responses can partially stabilize demand even under heightened risk. In the framework of Section 2, these dynamics correspond to an increase in expected activity in the "modal" state of the world even as the perceived probability of adverse tail states increases. That is exactly what generates risk-growth decoupling in which the distribution becomes more dispersed and more negatively skewed, raising risk premia, while the conditional expectation of medium-run activity can still rise due to mobilization and adaptation.

### 5.5.3 Implications for identification: cross-outcome coherence as evidence

The CLI result also strengthens identification through cross-outcome coherence. A standard concern in post-event macroeconomic analysis is that any single indicator can be contaminated by measurement issues, reporting delays, or idiosyncratic noise. Here, however, the sign pattern across outcomes is structured and theoretically consistent: long-horizon risk premia rise, household welfare beliefs deteriorate, and medium-run momentum improves. This configuration is difficult to reconcile with a single confounding shock unrelated to October 7, because most plausible macro confounders would tend to push these indicators in the same direction. Put differently, the CLI result is not an "oddity" to be explained away. It is part of the signature of a belief rupture. Specifically, he economy can move into a regime where growth momentum is supported or even strengthened by mobilization and reallocation while risk premia remain elevated because the left tail has thickened.

*5.5.4  Dynamics and the role of learning*

The sustained elevation in the CLI, alongside the hump-shaped pattern in sovereign spreads and the persistence in consumer confidence, suggests a specific time profile of belief updating. Medium-run momentum may stabilize or improve relatively quickly as policy and mobilization responses take effect and as production and adaptation mechanisms activate. In contrast, long-horizon tail risk may remain elevated because uncertainty about long-run geopolitical stability, fiscal costs, and institutional resilience resolves only slowly. Households' welfare beliefs can remain depressed due to salience, precautionary motives, and slow narrative updating even if momentum indicators improve. This temporal structure is consistent with a world in which learning operates at different speeds across horizons where medium-run activity becomes more predictable more quickly than long-horizon tail risk, and household beliefs adjust more slowly than market pricing.

*5.6  Venture Capital and the Global Financial Cycle*

Venture capital (VC) is a distinctive outcome because it is simultaneously forward-looking and globally priced. VC investment responds strongly to global discount rates, risk appetite, and exit conditions in international capital markets. It is therefore an ideal margin for separating local tail-risk repricing from global financial-cycle movements. In the conceptual framework, VC is governed less by domestic sovereign risk premia and more by global risk factors, particularly for an economy embedded in global technology finance networks. The matrix completion estimates suggest that VC investment in Israel does not decline relative to the donor pool in the post-treatment period. If anything, the estimated effect is modestly positive and statistically weakly significant. Therefore, the key empirical message is not that October 7 "increased VC," but that, conditional on the global cycle and relative to the donor pool, Israel does not exhibit an exceptional collapse in VC flows attributable to the shock.

This result is economically meaningful for three reasons. First, it implies that global venture financing conditions, already tightening in 2022-2023, dominate the level of VC activity in both Israel and the donor pool. In such an environment, a large domestic shock may not generate an additional measurable decline relative to comparators because the global financial cycle is the first-order driver. Second, it suggests that international investors may have viewed the shock as affecting sovereign risk and household welfare beliefs more than the expected private returns to high-growth tech projects, at least over the horizon captured by annual VC flows. Third, it is consistent with the risk-growth decoupling mechanism where long-horizon tail risk premia rise, yet high-growth sectors oriented toward global markets may remain resilient relative to global peers. This interpretation is reinforced by the fact that the VC result is estimated using annual data with a very short post-treatment sample. With only a small number of post-treatment observations, inference is necessarily limited. The absence of a strong negative relative effect should therefore be interpreted cautiously: it rules out a dramatic Israel-specific collapse beyond global VC contraction, but it does not imply that the domestic shock is irrelevant for VC at finer horizons or in specific segments of the funding market.

### 5.7  *Cross-Outcome Synthesis: A Horizon Map of Belief Updating*

At this stage, the results admit a coherent synthesis. The October 7 shock generates a structured reconfiguration of beliefs across horizons, which becomes visible precisely because the empirical design examines multiple outcomes rather than a single market series. Sovereign yields and sovereign spreads capture long-horizon repricing of tail risk. Their sustained elevation indicates that markets revised upward the perceived probability of adverse long-run states. Consumer confidence captures household welfare beliefs. Its sharp and persistent deterioration indicates a broad-based increase in perceived welfare risk and precautionary motives. Business tendency captures short-horizon firm expectations. Its lack of collapse, and possible improvement, is consistent with rapid adaptation and mobilization-driven demand at short horizons. The CLI captures medium-run momentum. Its strong rise indicates improved expectations about cyclical activity even in the presence of heightened tail risk. Venture capital captures exposure to the global financial cycle. Its absence of a large Israel-specific decline suggests that global conditions dominate, and that the domestic shock does not generate an additional collapse relative to peer economies at the annual frequency. This configuration constitutes what we refer to as belief-horizon heterogeneity, an extreme political rupture shifts the distribution of future states in a way that elevates long-horizon risk premia and depresses household welfare beliefs while leaving medium-run momentum and globally priced innovation finance comparatively resilient.

### 5.8  *Horizon Decomposition: Impact, Adjustment, and Persistence*

A core advantage of the event-time design is that it allows treatment effects to be decomposed into economically meaningful horizons rather than summarized by a single post-treatment average. This matters in our setting because October 7 is best understood as a rupture that triggers (i) an immediate reassessment of near-term disruption, (ii) an adjustment phase as information arrives and institutions respond, and (iii) a persistent shift in beliefs about long-run tail states. The dynamic treatment effect paths reported in the figures suggest that these phases differ sharply across outcomes. We operationalize this by partitioning the post-treatment period into horizon windows that correspond to the belief-horizon structure in Section 2. Let $H$ index horizons, and define three windows: an impact window $\mathcal{T}^I$ (short-run), an adjustment window $\mathcal{T}^A$ (medium-run), and a persistence window $\mathcal{T}^P$ (long-run). The horizon-specific effects are then:

$$ATET^H = \frac{1}{|\mathcal{T}^H|} \sum_{t \in \mathcal{T}^H} \tau_t, \quad H \in \{I, A, P\}$$

This decomposition does not merely sharpen presentation. It clarifies mechanism. For sovereign yields and especially the sovereign spread, the treatment effect exhibits a clear impact response followed by a build-up and then partial retracement. The persistence window remains decisively positive, indicating that the shock raised the long-horizon price of risk even after some resolution of acute uncertainty. This is exactly the pattern predicted by the belief-based mechanism in which agents initially overweight the left tail under extreme uncertainty and then partially revise as information arrives, without returning to the pre-shock baseline because the perceived distribution of future states has shifted.

For consumer confidence, the dynamic path is more consistent with a persistent negative shift and weaker mean reversion. This is also predicted by the model: household welfare beliefs are sticky and salient, and even if short-run adaptation occurs, perceived long-run welfare risk can remain elevated. The key conceptual point is that household confidence is not "corrected" by partial military or policy stabilization in the same way that market spreads can retrace; it embeds narrative and precautionary motives that update slowly. For the CLI, the discontinuity is sharp and the elevation is sustained. This suggests that the medium-run momentum channel activates quickly and remains active throughout the observed post-treatment horizon. In the model, this corresponds to the mobilization-and-reallocation component of the shock. Even as tail risk rises, the expected path of medium-run activity is revised upward, producing risk-growth decoupling. The horizon decomposition therefore delivers a crisp empirical message. The shock increases tail risk persistently, depresses welfare beliefs persistently, and increases medium-run momentum. The coexistence of these effects is precisely what distinguishes a belief rupture from standard macro disturbances.

## 5.9 Why the Joint Sign Pattern Rules Out Competing Explanations

A primary concern in any event-study of a major geopolitical rupture is that the estimated "effect" may in fact reflect other contemporaneous forces. Our identification strategy addresses this through imputation and placebo inference. But for a belief-based political rupture, there is an additional and powerful form of evidence of the joint pattern across outcomes. Competing macro explanations typically imply co-movement across these outcomes. The data instead show systematic divergence. A pure negative aggregate demand shock would be expected to reduce household confidence, reduce firm expectations, and reduce medium-run momentum, while its effect on long-term sovereign spreads would depend on policy response and the risk-free rate. The observed pattern contradicts this: consumer confidence falls, but the CLI rises sharply and business tendency does not collapse relative to the donor pool.

This is difficult to reconcile with a standard demand contraction. A pure negative supply shock would be expected to raise inflationary pressure and potentially raise nominal yields, while also reducing activity indicators and depressing confidence. Again, the observed rise in the CLI is not naturally implied by a negative supply shock, especially when the sovereign spread relative to the United States rises strongly, pointing toward country-specific risk rather than global inflation or policy. A purely monetary explanation, say, an Israel-specific tightening episode, would tend to raise yields and potentially depress confidence and activity measures. Yet the rise in the CLI and the resilience of short-horizon firm indicators are inconsistent with an unambiguous domestic tightening shock as the primary driver. Moreover, the use of the U.S. as a benchmark in the spread construction substantially reduces sensitivity to global monetary conditions.

A pure liquidity shock would be expected to manifest as a sharp spike in yields and spreads with rapid reversal once market functioning normalizes. Instead, the post-treatment spread is elevated for a sustained period, with only partial retracement. This persistence is inconsistent with purely technical dislocation and aligns with belief revision about long-horizon states. Finally, a pure fiscal expansion story might rationalize improved momentum indicators and possibly higher yields. But fiscal expansion alone cannot explain a sharp deterioration in consumer confidence alongside a rise in sovereign spreads relative to the United States, unless it also induces an upward revision in long-horizon tail risk. In other words, if fiscal expansion is part of the story, it operates through the very channel our framework emphasizes: belief updating about future fiscal capacity and tail states. The crucial point is that the joint sign pattern, higher sovereign risk premia, lower household welfare beliefs, and higher medium-run momentum, does not naturally emerge from standard one-factor macro shocks. It emerges naturally from a belief rupture that increases perceived tail risk while activating mobilization and adaptation forces that support medium-run activity. The multi-outcome evidence therefore serves as an additional layer of identification.

*5.10 Magnitudes and Economic Significance: Translating Treatment Effects into Economic Objects*

While the statistical evidence is strong, top-journal standards require a careful discussion of magnitudes. We therefore translate the key effects into economically interpretable objects, emphasizing ranges and avoiding overstated precision.

*5.10.1 Sovereign premia and the marginal cost of public capital*

The estimated increase in the sovereign spread relative to the United States is approximately 60-80 basis points in the baseline specifications. For sovereign debt pricing, such a repricing is economically meaningful. Even if applied only to marginal refinancing flows rather than the entire outstanding stock, it implies a sizable increase in annual debt service on newly issued long-term debt, with implications for fiscal space and the pricing of other domestic long-duration claims. More importantly, the sovereign spread is a sufficient statistic for the broader domestic cost of capital. Sovereign yields anchor domestic term premia, influence bank funding costs, and affect discount rates used by institutional investors. A persistent increase in the sovereign premium therefore implies a higher economy-wide hurdle rate, consistent with the belief-based tail-risk mechanism.

*5.10.2 Household confidence and welfare-belief shifts*

The consumer confidence growth effect is negative and persistent, with an estimated average decline of about 0.05 in the growth-rate metric. Although this cannot be mechanically translated into consumption without auxiliary elasticities, its economic significance lies in what it reveals about beliefs. It indicates a sustained deterioration in perceived welfare prospects, consistent with an increase in precautionary motives and in perceived downside risk. This is a channel through which an extreme political rupture can propagate beyond financial markets into household behavior, even in the absence of immediate collapse in medium-run momentum indicators.

*5.10.3 Medium-run momentum versus long-run risk: the macro-financial wedge*

The CLI effect, roughly +2 on the growth-rate scale, is large and persistent. The magnitude suggests a meaningful upward revision in medium-run momentum relative to what would have been predicted from Israel's pre-treatment comovement with advanced-economy donors. This supports the view that mobilization and reallocation mechanisms dominate at intermediate horizons.The coexistence of these magnitudes (i.e. a 60-80 bps rise in the sovereign premium), a sustained decline in household confidence, and a sizable increase in the CLI, is precisely what the model describes as a macro-financial wedge where risk premia rise even as momentum indicators improve. The shock therefore generates not a simple recessionary impulse, but a change in the pricing of states.

## 5.11  Relation to the literature

The results connect to, and help reconcile, several major strands of the modern macro-finance and political economy literature. First, the persistent rise in Israel's sovereign risk premium is consistent with theories in which political events alter equilibrium asset prices by changing expectations about policy regimes and the distribution of future states. In this view, political shocks do not operate solely through contemporaneous fundamentals; they operate through the discount rate and the pricing kernel. A prominent formulation of this mechanism is the political risk framework in which regime uncertainty and policy discontinuities are capitalized into risk premia and valuation ratios (Pastor and Veronesi 2012, 2013). Our findings provide direct empirical support for this channel using high-frequency sovereign pricing and an imputation-based counterfactual design that allows for heterogeneous exposure to global forces.

Second, the magnitude and persistence of the sovereign premium response is naturally interpreted through the lens of the rare-disaster and tail-risk pricing literature. In these models, small changes in the perceived probability of extreme adverse states can generate large movements in risk premia even when average growth expectations remain stable or improve (Barro 2006, Gabaix 2012). The empirical signature of such belief shifts is precisely the combination we document, a sustained increase in long-horizon risk pricing alongside improved medium-run momentum as captured by the CLI. This risk-growth decoupling is difficult to rationalize in one-factor frameworks but is a central implication of tail-risk pricing models.

Third, the divergence between household welfare beliefs (i.e. consumer confidence) and firm- or momentum-oriented indicators (i.e. business tendency, CLI) contributes to a growing literature emphasizing heterogeneous expectation formation and differential horizon updating across agents. Households update slowly and are strongly affected by salience, narratives, and precautionary motives, whereas markets and firms can incorporate information differently and focus on different horizons. Our multi-outcome design makes this heterogeneity visible and shows that it is not merely a behavioral nuance. It is a first-order feature of macro-financial adjustment after extreme political ruptures.

Fourth, methodologically, our design complements the synthetic control tradition by moving beyond static weight-based counterfactuals and exploiting low-rank structure in the outcome matrix. Recent advances emphasize that modern counterfactual estimators can be viewed as structured regularization problems, with different methods imposing different constraints (Athey et al. 2021; Arkhangelsky et al. 2021). We contribute by demonstrating that matrix completion, combined with rolling-window cross-validation and placebo-based inference, yields a credible identification strategy for a single treated unit under heterogeneous exposure to global forces, and that the resulting estimates are corroborated by SDID robustness. Overall, the results add a new empirical object to the literature: a high-frequency, multi-horizon mapping from an extreme political rupture to sovereign risk pricing, household welfare beliefs, medium-run momentum, and innovation finance. The joint pattern provides stronger evidence than any single series could deliver and offers a concrete empirical counterpart to models of belief-driven political risk.

## 5.12 *Robustness and Sensitivity: Credibility Beyond the Baseline*

The credibility of empirical estimates requires that the results be robust to reasonable perturbations of the estimation environment. We therefore conduct a battery of robustness checks that target the main threats to identification: (i) donor dependence, (ii) tuning-parameter sensitivity, (iii) pre-period instability, and (iv) estimator dependence. The aim is not to mechanically confirm significance but to establish that the central qualitative conclusions, the rise in the sovereign risk premium, the deterioration in household welfare beliefs, and the improvement in medium-run momentum, do not hinge on narrow modeling choices.

### 5.12.1 *Cross-estimator robustness: synthetic difference-in-differences*

We estimate synthetic difference-in-differences (SDID) for the same outcomes and treatment date. The SDID point estimates for the 10-year yield and the sovereign spread are close to the matrix completion estimates, on the order of 0.77-0.79 percentage points. While SDID inference is more conservative and confidence intervals are wider, the alignment of point estimates strongly suggests that the sovereign repricing is not an artifact of nuclear-norm regularization or latent-factor shrinkage. For consumer confidence and the CLI, SDID likewise confirms the direction of effects, even when precision varies due to the noisier nature of survey-based and composite indicators. The key takeaway is that two estimators with different identifying logic, one based on low-rank imputation and one based on optimized unit and time weighting, deliver consistent qualitative conclusions.

*5.12.2 Tuning-parameter sensitivity: the role of λ*

Because matrix completion relies on regularization, the choice of $\lambda$ is an important dimension of sensitivity. Our preferred $\lambda$ is selected via rolling-window cross-validation in the pre-treatment period, mimicking the true imputation task. We nonetheless report estimates under alternative values spanning more aggressive and more conservative shrinkage regimes. The sovereign yield and spread effects remain positive and economically large throughout, with magnitude varying in a disciplined way, stronger shrinkage yields larger estimated effects, reflecting the fact that less post-treatment variation is attributed to latent factors. This sensitivity is informative because it bounds the plausible effect size under alternative complexity assumptions, and it does not overturn the central conclusion of a substantial sovereign repricing.

*5.12.3 Donor pool sensitivity and leave-one-out tests*

We assess the dependence of results on donor composition in two ways. First, we perform leave-one-out exercises in which each donor country is removed in turn and the model is re-estimated. A credible design should not be driven by a single donor unit. Second, we estimate the model on donor subsets that exclude the most heavily weighted donors in SDID or the donors that appear most influential in fitting Israel's pre-treatment path. The sovereign-risk results remain qualitatively stable across these exercises, indicating that identification is not coming from a single "anchor" country but from a broader latent structure shared across the donor pool.

*5.12.4 Placebo inference: in-time and in-space placebos*

We conduct two placebo exercises. In-time placebos assign pseudo-treatment dates within the pre-treatment period for Israel. These tests are designed to detect spurious treatment effects arising from model misspecification, unmodeled trends, or overfitting. The pseudo-treatment effects are small and typically centered near zero, consistent with the notion that the estimator does not generate false discontinuities in periods without true treatment.

In-space placebos assign treatment to each donor country and compute placebo effects. This creates an empirical null distribution for the magnitude of estimated effects in units that were not exposed to the shock. Israel's estimated sovereign premium response lies in the extreme tail of this distribution, supporting the interpretation that the observed repricing is exceptional relative to untreated advanced economies.

*5.12.5 Window robustness and alternative post-period truncations*

Finally, we assess sensitivity to the sample window and post-treatment horizon. Because extreme political shocks can coincide with subsequent global events, we consider alternative truncations of the post-treatment window and verify that the core conclusions are not driven by any single month. The sovereign spread response remains positive and persistent under reasonable window choices. For the confidence and CLI measures, the direction of effects remains stable, though precision varies as expected when the post-treatment horizon is shortened.

*5.12.6 Synthetic difference-in-differences (SDID) estimates as a robustness check*

As a robustness check, we estimate treatment effects using synthetic difference-in-differences (SDID). SDID is particularly well suited to our setting—one treated unit, a long pre-treatment window, and potentially imperfect pre-treatment fit under any single method—because it combines two stabilizing elements. First, it constructs unit weights over the donor pool to reproduce the treated unit's pre-treatment path, as in synthetic control. Second, it constructs time weights over the pre-treatment period to remove residual imbalance through a difference-in-differences correction. The resulting estimand is a weighted DiD contrast that is less reliant on exact pre-treatment matching than classic synthetic control and less reliant on parallel trends than conventional DiD. Because SDID and matrix completion rely on fundamentally different counterfactual constructions, reweighted comparison versus low-rank imputation, agreement between them provides a strong validation that the baseline results are not estimator-specific.

Table 5 and Table 6 report the SDID estimates. For the sovereign-risk outcomes, SDID produces economically large positive effects of approximately 0.788 for the 10-year government bond yield and 0.771 for the yield spread relative to the United States. The corresponding confidence intervals are wide and include zero, and the permutation-based p-values are more conservative (on the order of 0.09-0.15 depending on the outcome and bootstrap scale). While these p-values do not uniformly meet conventional thresholds, the magnitude is strikingly close to the matrix-completion baseline, particularly under stronger shrinkage, and therefore corroborates the central conclusion that October 7 induced a substantial repricing of Israel's long-horizon sovereign risk. Figure 3 graphically reports SDID estimates in greater detail.

**Table 5**: Synthetic difference-in-differences estimates of the effect of October 7 on sovereign risk and bond yield, M1:2008-M8:2025

|  | 10-Year Government Bond Yield | | Sovereign Risk Premium (Bond Yield Spread) | |
|---|---|---|---|---|
|  | monthly | monthly | monthly | monthly |
|  | (1) | (2) | (3) | (4) |
|  | Small-scale (#100) | Large-scale (#1,000) | Small-scale (#100) | Large-scale (#1,000) |
| Panel A: Low hyper-space shrinkage parameter ($\lambda$=0.0001) | | | | |
| $\tau = \hat{\delta}_{Israel} - \sum_{i=1}^{N_{co}} \hat{\omega}_i \hat{\delta}_i$ | +0.788* | +0.788a | +0.771a | +0.771a |
|  | (0.468) | (0.547) | (0.533) | (0.516) |
| 95% Two-Tailed Empirical confidence intervals | {-0.129, 1.706} | {-0.284, 1.861} | {-0.267, 1.810} | {-0.40, 1.782} |
| Simulation-based empirical p-value (large-sample approximation) | 0.092 | 0.150 | 0.146 | 0.135 |
| Panel B: Composition of optimal country-level weights ($\Omega$) | | | | |
| Australia | 0.33 | | 0.35 | |
| Canada | 0 | | | |
| Denmark | 0.34 | | 0.39 | |
| Japan | 0 | | 0 | |
| South Korea | 0.19 | | 0.17 | |
| Norway | 0 | | | |

| | | |
|---|---|---|
| Sweden | 0.14 | 0.09 |
| United Kingdom | 0 | 0 |
| United States | 0 | 0 |

A distinctive advantage of SDID is transparency about the donor composition of the counterfactual. The estimated unit weights place most of the mass on a small subset of donors, especially Australia, Denmark, South Korea, and Sweden, with near-zero weights on the remaining countries. This concentration is informative: it indicates that the SDID counterfactual is not an undisciplined average of advanced economies but is formed from a coherent subset that best reproduces Israel's pre-treatment dynamics. It also suggests that the estimated sovereign repricing is not being driven by broad averaging across unrelated units; it persists when the counterfactual is built from the most predictive donors. SDID also largely preserves the cross-outcome structure emphasized in Section 5. For consumer confidence, SDID yields a negative effect (approximately -0.031) with a p-value around 0.056, consistent in sign with the matrix completion estimates but somewhat smaller in magnitude. For the composite leading indicator, SDID estimates a strongly positive effect of roughly +1.711, with a highly significant p-value (p-value = 0.001), confirming that the post-October 7 improvement in medium-run momentum is not an artifact of low-rank imputation.

For venture capital investment, SDID estimates an effect of about +0.296 (p-value = 0.011), closely aligned with the matrix completion results and consistent with the interpretation that global financial conditions dominate this margin at annual frequency. By contrast, SDID finds essentially no effect on business tendency, with a p-value close to one, indicating that this outcome is comparatively noisy and more sensitive to counterfactual construction. Taken together, the SDID results strengthen the paper's core claims in three ways. First, they corroborate the magnitude of the sovereign repricing under an estimator with a different identifying geometry. Second, they confirm the key cross-outcome signature of belief-horizon heterogeneity, particularly the sharp improvement in the CLI alongside higher sovereign premia and weaker household beliefs. Third, they help discipline interpretation by highlighting that business tendency is the least stable outcome across estimators, consistent with a higher-noise expectations measure rather than a central pillar of the empirical argument.

**Table 6**: Synthetic difference-in-differences estimates of the effect of October 7 on forward-looking expectations and beliefs, M1:2008-M8:2025

|  | Consumer confidence dynamics (growth rate) | Business tendency dynamics (growth rate) | Venture capital investment | Composite leading indicator |
|---|---|---|---|---|
|  | monthly | monthly | annual | monthly |
|  | (1) | (2) | (3) | (4) |
| Panel A: Synthetic difference-in-differences (SDID) estimates | | | | |
| $\hat{\tau} = \hat{\delta}_{Israel} - \sum_{i=1}^{N_{co}} \hat{\omega}_i \hat{\delta}_i$ | -0.031** | -0.028 | +0.296** | +1.711*** |
|  | (0.016) | (6.113) | (0.115) | (0.524) |
| 95% Empirical confidence intervals | {-0.063, 0.0007} | {-12.010, 11.954} | {0.066, 0.489} | {0.682, 2.738} |
| Simulation-based p-value | 0.056 | 0.996 | 0.011 | 0.001 |

Notes: This table reports synthetic difference-in-differences (SDID) estimates of the effect of the October 7, 2023 shock on forward-looking expectations and beliefs. Each column reports the average post-treatment treatment effect for Israel relative to a weighted combination of donor countries, where the weights are chosen to optimally balance pre-treatment outcomes across both units and time. The SDID estimator combines unit-level reweighting, as in synthetic control, with time-level reweighting, as in difference-in-differences, and allows for heterogeneous treatment timing and factor structures. Standard errors, confidence intervals, and p-values on null hypothesis are obtained using simulation-based inference based on repeated placebo assignments following Arkhangelsky et al. (2021). Reported confidence intervals are empirical, finite-sample intervals. Asterisks denote statistical significance at the 10% (*), 5% (), and 1% (*) levels.

**Figure 3**: Synthetic difference-in-differences estimates of the effect of October 7 on sovereign risk and bond yield, M1:2008-M8:2025

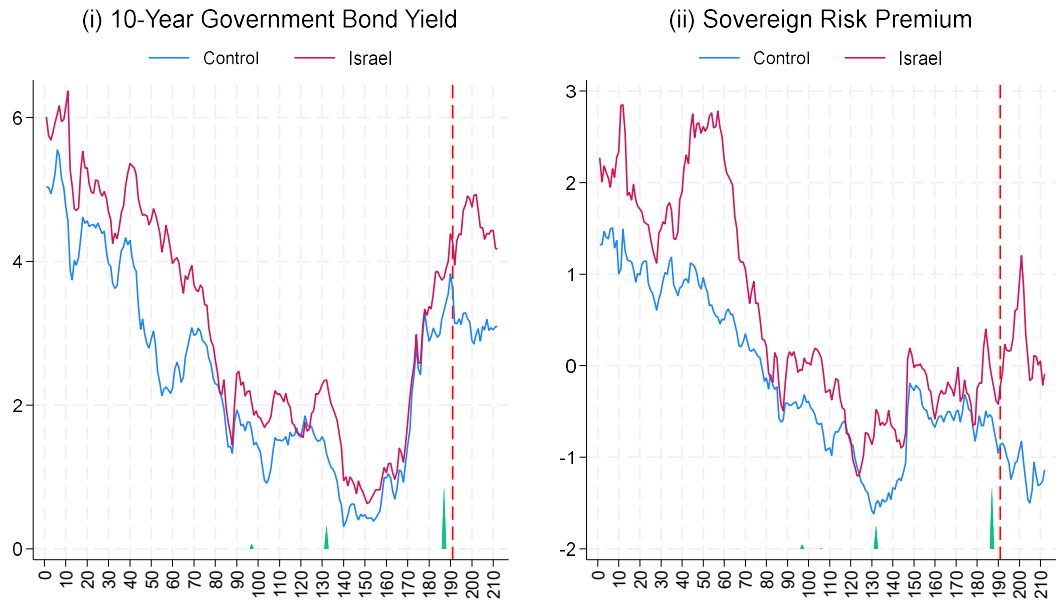

## 6 Discussion and Policy Implications

This section interprets the results through the lens of the framework developed in Section 2 and draws policy-relevant implications. The aim is not to claim that a single event "determines" macroeconomic outcomes, but to identify what October 7 reveals about the pricing of tail risk, the formation of beliefs, and the transmission of extreme political ruptures into the cost of capital and the real economy.

### 7.1 October 7 as a Belief Rupture: What the Evidence Implies

The joint pattern across outcomes is most naturally interpreted as a belief rupture rather than as a standard macroeconomic disturbance. Sovereign yields and, especially, the sovereign spread relative to the United States rise sharply and remain elevated, implying a persistent increase in the compensation investors require to hold long-duration Israeli sovereign risk. Household welfare beliefs deteriorate substantially, consistent with a durable upward revision in perceived downside risk. Yet medium-run momentum, as captured by the OECD CLI, improves strongly, and venture capital does not collapse relative to the donor pool at annual frequency. The economy therefore exhibits risk-growth decoupling where tail risk premia rise even as medium-horizon momentum indicators strengthen. This configuration is informative precisely because it is not mechanically implied by common one-factor narratives. A pure demand contraction would depress both confidence and momentum. A purely technical liquidity event would not sustain a country-specific spread premium. A purely global risk-off episode would be largely absorbed by time effects and would not produce a distinctive Israel-donor pool divergence. The evidence instead points to an economy that adapts along operational and medium-horizon dimensions while the perceived distribution of long-run states becomes more negatively skewed.

### 7.2 *Sovereign Premia as a Constraint on Fiscal Space and Public Risk-Bearing*

The increase in Israel's sovereign risk premium is economically important even if one abstracts from precise fiscal arithmetic. Sovereign yields anchor the discount rate for long-horizon public projects and affect the pricing of domestic long-duration assets more broadly. A persistent increase in the sovereign spread therefore acts as a tax on public risk-bearing. It raises the hurdle rate for government borrowing and compresses fiscal space precisely at a time when security-related spending needs and social insurance demands may rise. From a policy perspective, this is the key implication of the sovereign risk result: extreme political ruptures can impose a persistent "tail-risk wedge" on public finance. The wedge is not necessarily proportional to immediate economic disruption, it reflects long-horizon uncertainty about future states. This suggests that stabilization policy after such shocks cannot be assessed solely by short-run output or employment performance. A country can appear resilient in medium-run momentum indicators and yet face a higher long-run cost of public capital.

### 7.3 *Household Beliefs and the Political Economy of Welfare Risk*

The consumer confidence evidence highlights a second channel with direct policy relevance: households internalize extreme political shocks as welfare risk, and these beliefs can remain depressed even when real activity indicators stabilize. In the framework of Section 2, household belief updating affects consumption, precautionary saving, and labor market behavior. Persistently depressed confidence can therefore dampen private demand and amplify inequality in exposure to uncertainty, because the capacity to smooth shocks differs across households. This has two implications. First, post-shock stabilization policy may need to target not only liquidity and credit provision, but also the distribution of welfare risk, including income insurance and credible commitments about future policy. Second, the persistence of household belief responses suggests that communication policy and institutional credibility can be macroeconomically relevant: when households perceive that the probability mass in adverse states has increased, purely technical macro stabilization may be insufficient to restore confidence.

### 7.4 *Why Momentum Can Rise When Tail Risk Increases*

Perhaps the most conceptually important implication of the results is that momentum and tail risk can move in opposite directions. The strong positive response of the CLI indicates that medium-run activity expectations improved relative to the counterfactual, consistent with mobilization, reallocation, and policy response. This can coexist with a higher sovereign premium because the premium is determined by the left tail of the distribution, not by the modal forecast. This matters for policy evaluation. Governments and observers often infer "recovery" from improvements in momentum indicators. Our results caution against equating momentum with restored stability. An economy can generate strong medium-horizon momentum, through reallocation, fiscal support, or wartime mobilization, while the long-horizon price of risk remains elevated. From the perspective of long-run investment and public finance, the latter may be the binding constraint.

### 7.5 *Private Capital, Innovation Finance, and External Pricing*

The venture capital results suggest that Israel's innovation finance is priced to a substantial extent on global terms. The absence of an exceptional Israel-specific collapse relative to the donor pool at annual frequency is consistent with the dominance of the global financial cycle in venture funding and with the international nature of Israel's tech ecosystem. This does not imply that domestic shocks are irrelevant for innovation, but it suggests that their impact may operate through composition (which firms get funded, at what terms, and with what relocation or governance responses) rather than through aggregate annual volume alone. For policy, the implication is nuanced. On the one hand, integration into global capital markets can buffer innovation finance against domestic shocks. On the other hand, a persistent increase in sovereign risk premia could still transmit into the private sector through domestic credit conditions, exchange-rate risk, and the valuation of long-duration cash flows. An important extension is therefore to examine whether the cost of private capital rises even when volumes remain resilient—for example via spreads on corporate debt, equity risk premia, or the terms of late-stage financing rounds.

## 7.6  External Validity: What This Case Teaches About Political Ruptures and the Cost of Capital

Although this paper studies a specific event, the results contribute to a broader understanding of how extreme political ruptures affect the cost of capital. The main lesson is not that "wars raise yields." Rather, the evidence suggests that rare political shocks operate by shifting beliefs about long-horizon states, and that these belief shifts can persist even as short- and medium-horizon indicators stabilize. This mechanism is likely to apply in other settings where geopolitical risk, institutional fragility, or regime uncertainty thickens the left tail of the distribution. The methodological implication is also general: in macro-financial settings with heterogeneous global exposure, designs that flexibly model latent factors such as matrix completion can deliver credible counterfactuals when combined with disciplined regularization and placebo inference. The substantive implication is that the cost of capital is shaped not only by realized fundamentals, but by the perceived distribution of future states.

## 7.7  Policy Takeaways

The findings suggest four policy-relevant takeaways. First, extreme political ruptures can impose a persistent increase in the sovereign risk premium, tightening fiscal space even if medium-run activity rebounds. Second, households internalize such shocks as durable welfare risk, which can suppress confidence and potentially weigh on private demand. Third, improvements in momentum indicators should not be interpreted as evidence that long-horizon risk has normalized, risk-growth decoupling can persist. Fourth, global integration of innovation finance may buffer aggregate venture investment volumes, but it does not eliminate the possibility that the cost of private capital rises or that the composition of investment changes.

## 8  Conclusion

This paper studies how an extreme political rupture reshapes the cost of capital, expectations, and forward-looking macro-financial indicators, using the October 7, 2023 shock as a sharply dated treatment event for Israel. Motivated by a belief-based framework in which rare political shocks reweight the distribution of future states, thickening the left tail and shifting long-horizon risk premia, we construct a transparent counterfactual using modern panel imputation methods and high-frequency data.

Empirically, we assemble a monthly panel spanning M1:2008-M8:2025 for Israel and an advanced-economy donor pool, combining sovereign yields and sovereign spreads from Bloomberg with OECD measures of consumer confidence, business tendency, the composite leading indicator, and venture capital investment. The frequency and breadth of outcomes allow us to map the effects of the shock across horizons, specifically, long-horizon risk pricing, household welfare beliefs, firm-side expectations, medium-run momentum, and innovation finance. Our identification strategy treats causal inference as a structured missing-data problem. We estimate Israel's post-treatment counterfactual path using the matrix completion estimator with unit and time fixed effects and nuclear-norm regularization, selecting the regularization parameter via rolling-window cross-validation on the pre-treatment block. Inference is conducted using both in-time and in-space placebo designs. The results are corroborated using synthetic difference-in-differences, an estimator with a distinct identifying geometry based on joint unit and time reweighting.

Three headline findings emerge. First, October 7 induces a large and persistent repricing of Israel's long-horizon sovereign risk. Both 10-year government bond yields and yield spreads relative to the United States rise sharply after the shock and remain elevated, consistent with a sustained increase in perceived tail risk rather than a purely global interest-rate movement or transitory liquidity disruption. Second, household welfare beliefs deteriorate sharply and persistently, as captured by a large decline in consumer confidence growth, indicating that the shock shifts households' perceived distribution of future welfare states in a durable way. Third, medium-run momentum improves relative to the counterfactual, the composite leading indicator rises strongly, suggesting rapid adaptation and mobilization effects that support activity expectations even as long-horizon risk premia remain elevated. Venture capital investment does not exhibit an exceptional collapse relative to the donor pool at annual frequency, consistent with the dominance of the global financial cycle in that margin and with the possibility that the shock operates through risk premia and composition rather than aggregate volume alone.

Taken together, these findings document a clear case of risk-growth decoupling following an extreme political rupture. The economy can exhibit resilience in medium-run momentum measures while the long-horizon cost of sovereign capital rises and households internalize the shock as persistent welfare risk. This pattern is difficult to reconcile with standard one-factor macro narratives and is naturally explained by belief-driven repricing of tail states. The results have several implications. For policy, they suggest that post-shock stabilization should be evaluated not only through short-run output and momentum, but through the evolution of long-horizon risk premia and household welfare beliefs, which govern fiscal space and private precautionary behavior. For research, they highlight the value of multi-outcome designs and high-frequency data for distinguishing between short-run adaptation and long-horizon repricing, and they highlight the usefulness of low-rank imputation methods, disciplined by cross-validation and placebo inference, in settings with heterogeneous global exposure and a single treated unit.

Several extensions are natural. First, one may examine whether the rise in sovereign risk premia transmits into private borrowing costs, equity risk premia, or the terms (rather than volumes) of innovation finance. Second, a richer heterogeneity analysis could study sectoral responses and the distribution of belief shifts across households. Third, future work could compare the October 7 episode to other extreme political ruptures to quantify how institutional credibility, fiscal capacity, and geopolitical embeddedness shape the persistence of tail-risk repricing. More broadly, the evidence reinforces a central lesson of modern macro-finance: the cost of capital is shaped not only by realized fundamentals, but by how political events alter the perceived distribution of future states.